\documentclass{aa}
\usepackage[utf8]{inputenc}
\usepackage[T1]{fontenc}
\usepackage{natbib}
\usepackage{amsmath}
\usepackage{amssymb}
\usepackage{geometry}
\usepackage{graphicx}
\usepackage{bm}
\usepackage{indentfirst}
\usepackage{hyperref}
\usepackage{subfig}
\usepackage{xcolor}
\usepackage{txfonts}
\usepackage{ulem}
\bibpunct{(}{)}{;}{a}{}{,}
\usepackage{cleveref}
\usepackage{braket}
\usepackage{relsize}
\usepackage{stackengine,wasysym}
\usepackage{scalerel}

\renewcommand*\d{\mathop{}\!\mathrm{d}}
\newcommand\bigsum[1]{{\mathlarger{\sum}}_{#1=1}^N}
\newcommand\xiosc{\bm{\xi}_\mathrm{osc}}
\newcommand\xiosci[1]{\xi_\mathrm{osc,#1}}
\newcommand\uosc{\mathbf{u}_\mathrm{osc}}
\newcommand\uosci[1]{u_\mathrm{osc,#1}}
\newcommand\ut{\mathbf{u}_t}
\newcommand\uti[1]{u_{t,#1}}
\newcommand\reallywidetilde[1]{\ThisStyle{%
  \setbox0=\hbox{$\SavedStyle#1$}%
  \stackengine{-.1\LMpt}{$\SavedStyle#1$}{%
    \stretchto{\scaleto{\SavedStyle\mkern.2mu\AC}{.5150\wd0}}{.6\ht0}%
  }{O}{c}{F}{T}{S}%
}}

\begin{document}

\title{Coupling between turbulence and solar-like oscillations: A combined Lagrangian PDF/SPH approach}
\subtitle{I -- The stochastic wave equation}

\author{J. Philidet\inst{\ref{inst1}} \and K. Belkacem\inst{\ref{inst1}} \and M.-J. Goupil\inst{\ref{inst1}}}

\institute{LESIA, Observatoire de Paris, PSL Research University, CNRS, Universit\'e Pierre et Marie Curie, Universit\'e Paris Diderot, 92195 Meudon, France \label{inst1}}

\abstract
{The development of space-borne missions such as CoRoT and \textit{Kepler} now provides us with numerous and precise asteroseismic measurements  that allow us to put better constraints on our theoretical knowledge of the physics of stellar interiors. In order to utilise the full potential of these measurements, however, we need a better theoretical understanding of the coupling between stellar oscillations and turbulent convection.}
{The aim of this series of papers is  to build a new formalism specifically tailored to study the impact of turbulence on the global modes of oscillation in solar-like stars. In building this formalism, we circumvent some fundamental limitations inherent to the more traditional approaches, in particular the need for separate equations for turbulence and oscillations, and the reduction of the turbulent cascade to a unique length and  timescale. In this first paper we   derive a linear wave equation that directly and consistently contains the turbulence as an input to the model, and therefore naturally contains the information on the coupling between the turbulence and the modes through the stochasticity of the equations.}
{We use a Lagrangian stochastic model of turbulence based on probability density function methods to describe the evolution of the properties of individual fluid particles through stochastic differential equations. We then transcribe these stochastic differential equations from a Lagrangian frame to a Eulerian frame  more adapted to the analysis of stellar oscillations. We combine this method with smoothed particle hydrodynamics, where all the mean fields appearing in the Lagrangian stochastic model are estimated directly from the set of fluid particles themselves, through the use of a weighting kernel function allowing to filter the particles present in any given vicinity. The resulting stochastic differential equations on Eulerian variables are then linearised. As a first step the gas is considered to follow a polytropic relation, and the turbulence is assumed anelastic.}
{We obtain a stochastic linear wave equation governing the time evolution of the relevant wave variables, while at the same time containing the effect of turbulence. The wave equation generalises the classical, unperturbed propagation of acoustic waves in a stratified medium (which reduces to the exact deterministic wave equation in the absence of turbulence) to a form that, by construction, accounts for the impact of turbulence on the mode in a consistent way. The effect of turbulence consists of a non-homogeneous forcing term, responsible for the stochastic driving of the mode, and a stochastic perturbation to the homogeneous part of the wave equation, responsible for both the damping of the mode and the modal surface effects.}
{The stochastic wave equation obtained here represents our baseline framework to properly infer properties of turbulence-oscillation coupling, and can therefore be used to constrain the properties of the turbulence itself with the help of asteroseismic observations. This will be the subject of the rest of the papers in  this series.}


\maketitle

\section{Introduction}

Solar-like oscillations are coupled with turbulent convection in a complex manner, especially in the highly turbulent subsurface layers of the star \citep[see][for a review]{samadiB15,houdekMADreview}. This coupling impacts the behaviour of the modes in several major ways. One of the most prominent effects concerns mode frequencies, and explains in a large part the systematic discrepancy between the theoretical and observed $p$-mode frequencies \citep{dziembowski88,jcd96,rosenthal99}. The variety of physical processes responsible for the impact of turbulent convection on $p$-mode frequencies is collectively referred to as `surface effects'. These surface effects constitute a major obstacle preventing us from using the full potential of modal frequencies for an accurate probing of stellar interiors or for a precise inference of stellar global parameters.

Many efforts have thus been devoted to the correction of surface effects, either from theoretical modelling \citep[e.g.][]{gabriel75,balmforth92b,houdek96thesis,rosenthal99,grigahcene05} or through empirical formulae \citep[e.g.][]{kjeldsen08,jcd12,ballG14,sonoi15}. Some aspects, however, are very complicated to model, and existing models make use of assumptions that can barely be justified, if at all. For instance, turbulent pressure modulations are usually described in the Gas-$\Gamma_1$ (GGM) or reduced-$\Gamma_1$ (RGM) approximations \citep{rosenthal99}, which amounts to neglecting the effects of turbulent dissipation and buoyancy on the mode \citep{belkacem21}. Another problem is the use of time-dependent mixing-length formalisms \citep{unno67,gough77a} to account for modal surface effects \citep[e.g.][]{gabriel75,houdek96thesis,grigahcene05,sonoi17,houdek17,houdek19}.  While useful for the bulk of the convective region, the mixing-length hypothesis is no longer valid in the superadiabatic region just beneath the surface of the star, as shown by 3D hydrodynamic simulations of stellar atmospheres \citep[see][for a review]{nordlund09review}. Finally, such formalisms require that the oscillations be separated from the convective motions, thus yielding separate equations. This is done either by assuming a cut-off in wavelength space, with oscillations having much shorter wavelengths than turbulent convection \citep{grigahcene05}, or by using 3D hydrodynamic simulations and separating the oscillations from convection though horizontal averaging \citep{nordlundS01}. The necessity to separate the equations for oscillations and convection is fundamentally problematic as there is no rigorous way to disentangle the two components, mainly because, in solar-type stars, they have the same characteristic lengths and timescales \citep{samadiB15}. This is even truer if one wishes to model their mutual coupling.

Turbulent convection also has a crucial impact on the energetic aspects of solar-like oscillations. Solar-like $p$-modes are stochastically excited and damped by turbulent convection at the top of the convective zone. As such, understanding the energetic processes pertaining to the oscillations leads to better constraints on the highly turbulent layers located beneath the surface of these stars. Many theoretical modelling efforts were deployed on the subject of mode driving \citep[e.g.][]{goldreichK77a,goldreichK77b,balmforth92a,balmforth92c,samadiG01,chaplin05,samadi05,samadi06,belkacem06,belkacem08,belkacem10}, as well as mode damping \citep[e.g.][]{goldreichK91,balmforth92a,grigahcene05,dupret06,belkacem12}. The fact that these energetic processes take place in the superadiabatic region, however, makes any predictive model extremely complicated to design as this requires a time-dependent non-adiabatic turbulent convection model able to include the oscillations. Subsequently, modelling attempts have focused on the use of mixing-length formalisms to account for mode damping. This approach, however, presents the considerable disadvantage of reducing the turbulent cascade to a single length scale, and is therefore unable to correctly account for the contribution of turbulent dissipation or turbulent pressure to mode damping. Alternative approaches have been followed in an attempt to go beyond the mixing-length hypothesis, either through a Reynolds stress model \citep{xiong00} or through the use of 3D hydrodynamic simulations of stellar atmospheres to directly measure mode linewidths \citep{belkacem19,zhou20}.

These traditional approaches therefore show some fundamental limitations, which prevent them from being able to fully describe the interaction between turbulent convection and oscillations, whether it be to explain the surface effects on mode frequency or the energetic aspects of global modes of oscillation regarding their driving and damping physical processes. Among these limitations, we can include the following.

These approaches require that the turbulent convection and the oscillations be separated into two distinct sets of equations from the start. This is usually justified either by a separation of spatial scales or timescales, or else by performing some averaging process designed to separate an average component from a fluctuating component. The necessity of artificially separating these two intertwined phenomena from the outset is problematic when it comes to modelling their coupling.

Most of these approaches are based on a time-dependent mixing-length formalism, which oversimplifies the behaviour of convection in the superadiabatic region. In addition to poorly describing the structure of the convective motions close to the surface, it reduces the turbulent cascade to a single characteristic length scale, thus only offering a crude understanding of turbulent dissipation, a phenomenon deemed crucial to turbulence--oscillation coupling.

In these approaches, designing a closure relation for the model equations is a complicated process. In particular, it is very difficult to properly relate the chosen closure to the underlying physical assumptions. This is illustrated, for instance, by the wealth of free parameters that need to be adjusted in approaches based on mixing length theory (MLT) or in Reynolds stress approaches where higher-order moments need to be closed at the mean flow level. Approaches based on 3D simulations are not spared, as is illustrated by the need to rely on assumptions like the GGM or RGM, which are not clearly physically grounded.

The multiple free parameters needed in these approaches, and the fact that they are not easily constrained physically, also presents the distinct disadvantage of robbing these models from their predictive power. This becomes problematic when, for example, mode damping rates are used in scaling relations for seismic diagnosis purposes \citep{houdek99,chaplin09,baudin11,belkacem12}. The exponent in these scaling relations is difficult to determine, and varies substantially across the Hertzsprung--Russell diagram. Being able to predict the damping rates of stars with different global parameters would go a long way towards a more effective use of this quantity in such scaling relations.

Model parameters for the surface effects on the one hand, and mode damping rates on the other are usually constrained by completely separate adjustment procedures. This is also problematic, as these two quantities are closely related, and are actually just two sides of the same phenomenon:  the real and imaginary part of the turbulence-induced shift in the complex eigenfrequency of the modes.

These limitations form substantial hurdles towards a correct turbulence--oscillation coupling model, and circumventing them requires going beyond the methods presented above. Therefore, this series of papers follows a completely different approach. More precisely, the fundamental motivation behind this work is to provide  a method that 1) does not initially rely on a separation between convection equations and oscillation equations, but instead encompasses both components at the same time, and therefore naturally contains their coupling; 2) avoids the reduction of length scales in the problem to a unique scale, but instead accounts for the full description of the turbulent cascade; 3) simultaneously describes all effects of turbulent convection on mode properties, including the surface effects and the energetic aspects pertaining to mode driving and damping, in a single consistent framework; and 4) includes the properties of turbulence in such a way that they can be easily related to the observed properties of the modes. 

In this paper we therefore build a formalism for the modelling of turbulence--oscillation coupling, which is based on probability density function (PDF) models of turbulence \citep[e.g.][]{obrien80book,pope85,popeC90,pope91,pope94,vanslooten98}. The quantity we model is the PDF associated with the random flow variables, whose evolution follows a transport equation that takes the form of a Fokker-Planck equation \citep{gardiner85book}. Because the Fokker-Planck equation is impractical to handle both analytically and numerically, the PDF is usually represented by a set of fluid particles constituting the flow. The properties of the particles evolve according to stochastic differential equations, and are then used to reconstruct any given statistics of the flow. This is at the heart of Lagrangian stochastic models of turbulence, which have been used extensively by the fluid dynamics community, first for incompressible flows \citep[e.g.][]{pope81,anand89,haworth91,roekaerts91}, and then for compressible flows as well \citep[e.g.][]{hsu94,delarueP97,weltonP97,welton98,dasD05,bakosiR11}. In this series of papers we present a general way of using such a Lagrangian stochastic model of turbulence to derive a linear stochastic wave equation applicable to the stellar context. The wave equation is designed to govern the physics of the modes, while simultaneously and consistently encompassing the impact of turbulent convection thereon. This first paper describes how the linear stochastic wave equation is obtained. A subsequent paper will present how this wave equation can be used to simultaneously model the turbulence-induced surface effects, as well as the stochastic driving and damping of the modes by turbulent convection.

This paper is structured as follows. In Section \ref{sec:stoch-model} we introduce the stochastic model of turbulence that we will use throughout this study in terms of Lagrangian variables. Then we carry out a variable transformation to obtain stochastic equations on Eulerian quantities, which are more suitable for stellar oscillation analysis. In Section \ref{sec:waveeq} we linearise the Eulerian stochastic equations to obtain a  linear stochastic wave equation, and then discuss how it relates to other more familiar forms of the wave equation found in the literature and obtained through more traditional methods. Finally, in Section \ref{sec:discussion} we return to the various simplifications and approximations adopted in the present derivation, and what they entail as regards the resulting properties of turbulence-oscillation coupling. Conclusions are drawn in Section \ref{sec:conclusion}.

\begin{table}
    \centering
    \begin{tabular}{ll}
        Symbol & Definition \\
        \hline \\
        $f_0$ & Time-averaged equilibrium value of the quantity $f$ \\
        $\overline{f}$ & Instantaneous Reynolds average of $f$ \\
        $\widetilde{f}$ & Instantaneous Favre average of $f$ \\
        $\langle f \rangle_L$ & Lagrangian-mean of $f$ \\
        $f_1$ & Fluctuation of $f$ around $f_0$ \\
        $f'$ & Fluctuation of $f$ around $\overline{f}$ \\
        $f''$ & Fluctuation of $f$ around $\widetilde{f}$ \\
        $a_i$, $b_{ij}$ & Drift and diffusion coefficients in velocity stochastic \\
        & differential equation (SDE) \\
        $C_0$ & Kolmogorov constant \\
        $c_0$ & Equilibrium sound speed \\
        $\epsilon$ & Turbulent dissipation rate \\
        $\eta_i$ & Time derivative of Wiener process \\
        $g_i$ & Gravitational acceleration \\
        $G_{ij}$ & Drift tensor in velocity SDE \\
        $\gamma$ & Polytropic exponent \\
        $\Gamma_1$ & First adiabatic exponent \\
        $k$ & Turbulent kinetic energy \\
        $K$ & Kernel weighting function in SPH formalism \\
        $p$ & Gas pressure \\
        $\rho$ & Gas density \\
        $u_i$ & Flow velocity in Eulerian frame \\
        $u_i^\star$ & Fluid particle velocity in Lagrangian frame \\
        $u_{i,t}$ & Turbulent part of $u_i$ \\
        $u_{i,\text{osc}}$ & Oscillatory part of $u_i$ \\
        $W_i$ & Wiener process \\
        $x_i$ & Eulerian average position of fluid particle \\
        & (used as Eulerian space variable) \\
        $x_i^\star$ & Fluid particle position in Lagrangian frame \\
        $X_i$ & Instantaneous fluid particle position, as a \\
        & function of Eulerian average position \\
        $\xi_i$ & Fluctuation of $X_i$ around $x_i$ \\
        $\xi_{i,t}$ & Turbulent part of $\xi_i$ \\
        $\xi_\text{i,osc}$ & Oscillatory part of $\xi_i$ \\
        $\omega_t$ & Turbulent frequency
    \end{tabular}
    \caption{Glossary of the notations used in this paper.}
    \label{tab:glossary}
\end{table}

\section{Stochastic model of turbulence\label{sec:stoch-model}}

In MLT formalisms the modelled quantities pertain to the mean flow (e.g. mean density, velocity, entropy), and the second-order moments appearing in the mean equations must be expressed in terms of the mean flow in order to close the system. In Reynolds stress formalisms the closure at second-order level is replaced by equations on second-order moments, where third-order moments must be similarly closed. These moments are all defined as ensemble averages of stochastic processes such as flow velocity and entropy.  The core idea behind PDF models is to replace these numerous equations on various statistical moments of turbulent quantities by a single equation on the PDF of these quantities, in the form of a Fokker-Planck equation. These models present several advantages, which are of special interest given the issues raised in the previous section. By nature, the modelled PDF contains all the required statistical information on the flow, which includes both the turbulent convection and the oscillating modes. As such, this type of model is perfectly suited for the study of turbulence--oscillation coupling. In addition, all the usual quantities can be obtained from the PDF.

However, the direct modelling of the PDF, using its time evolution equation, can quickly become very cumbersome. The reason is that the PDF is a function not only of space and time, but also of each of the turbulent quantities used to represent the flow (starting with the three components of the velocity and the entropy). This makes the PDF equation computationally heavy to integrate, and quite impractical to handle analytically. This is why PDF models are often implemented in a Lagrangian particle framework, where the flow is no longer represented by a set of Eulerian, grid-based fluid quantities, but rather by a set of individual fluid particles whose properties (including their position) are tracked over time. Using Monte Carlo methods, the flow PDF can be reconstructed directly from the set of particles, so that the set contains the exact same statistical information as the PDF itself. In order to represent the turbulent nature of the flow, and to model the PDF accurately, particle properties must evolve according to stochastic differential equations rather than ordinary ones. Therefore, PDF models of turbulence go hand in hand with the implementation of Lagrangian stochastic methods, primarily because it makes their numerical integration much easier and more tractable. In this section we introduce the Lagrangian stochastic model, and we present how it can be rearranged to yield stochastic differential equations for Eulerian quantities instead.

We note that this paper aims to show that the method  we present is relevant to the study of turbulence--oscillation coupling, and therefore serves as a proof of concept for this approach. As such, we do not claim to use the most realistic turbulence model possible, but rather we wish to limit the level of complexity so that the basics of this method may be understood in the most efficient way. We leave the use of a more realistic turbulence model for a later paper.

\subsection{Lagrangian description: The generalised Langevin model\label{sec:GLM}}

We consider the simplified case of an adiabatic\footnote{In the vocabulary of asteroseismology, the term `adiabatic' can sometimes be used to express the absence of energy transfer between the oscillations and the background. We insist that this is not the case here. This term is meant to apply to the thermodynamic transformations undergone by the flow, not to the oscillations. In particular, the background can still inject energy into the modes  or take energy from them, allowing  the modes to be driven and damped.} flow, in the sense that we do not include an energy equation in the system, and instead  adopt a polytropic relation between the pressure and the density of the gas. In terms of Eulerian transport equations that would mean only considering the density, the mean velocity, and the Reynolds stress tensor as relevant fluid quantities, with the mean pressure being given, for instance, by the ideal gas law. In the framework of a Lagrangian stochastic model, however, that means that the only fluid particle properties whose evolution we need to put into equations are their position and velocity.

The equation for the particle position is derived by stating that it must evolve according to its own velocity. It reads
\begin{equation}
    \d x_i^\star = u_i^\star \d t~,
    \label{eq:stochx}
\end{equation}
where $\mathbf{x}^\star$ and $\mathbf{u}^\star$ are the position and velocity of the fluid particle, which only depend on the time variable (as well as the initial state). In general, in the following the notation $^\star$ will denote a stochastic variable. In order to account for the turbulent nature of the flow, the equation on velocity must take the form of a stochastic differential equation (SDE), instead of an ordinary one. In its most general form, an SDE takes the form \citep[][Chap. 3]{gardiner85book}
\begin{equation}
     \d u_i^\star = a_i(\mathbf{x}^\star, \mathbf{u}^\star, t) \d t + b_{ij}(\mathbf{x}^\star, \mathbf{u}^\star, t) \d W_j~,
    \label{eq:stochu}
\end{equation}
where we use the Einstein convention on repeated indices, $a_i$ and $b_{ij}$ are functions of the particle properties (and time), and $\mathbf{W}(t)$ is an isotropic Wiener process. The last is a stochastic process (i.e. a random variable whose statistical properties depend on time) whose PDF at any given time $t$ is Gaussian, and which verifies
\begin{align}
    & \overline{\mathbf{W}(t)} = \mathbf{0}~, \\
    & \overline{W_i(t') W_j(t)} = (t' - t) ~ \delta_{ij}~, \label{eq:W-var}
\end{align}
where $\delta_{ij}$ is the Kronecker symbol and the notation $\overline{\vphantom{u}~.~}$ refers to an ensemble average. We note that this is not a simplifying assumption regarding the stochastic part of the SDE, but rather a very general property, which is necessary for the resulting particle trajectory in phase-space to be continuous in time \citep{gardiner85book}. In terms of dimension, the drift vector $a_i$ is an acceleration, while $\mathbf{W}$ is the square root of a time, and the diffusion tensor $b_{ij}$ is a velocity divided by the square root of a time.

On the right-hand side of Eq. \eqref{eq:stochu} the first term corresponds to the deterministic part of the force exerted on the fluid particle, while the randomness of the equation is only brought about by the second term. Physically, the stochastic part of Eq. \eqref{eq:stochu} stems from the fluctuating components of both the pressure and viscous stress forces, which in turn are brought about by the underlying highly fluctuating turbulent velocity field. An illuminating analogy to consider is Brownian motion, which can also be described by means of Eq. \eqref{eq:stochu}, and where the stochastic part describes the random collision undergone by the colloidal particle from the water molecules. In the vocabulary of stochastic processes the function $a_i(\mathbf{x}^\star, \mathbf{u}^\star, t)$ is the $i$-th component of the drift vector, while $b_{ij}(\mathbf{x}^\star, \mathbf{u}^\star, t)$ is the $i,j$-component of the diffusion tensor. In order to close the system,  an explicit expression is needed for these two coefficients.

The specification of the drift and diffusion terms in Eq. \eqref{eq:stochu} is the subject of an abundant amount of literature on turbulence modelling \citep[see][for a review]{heinz04book}. It has long been recognised that, in order to be consistent with the Kolmogorov hypotheses, both original \citep{kolmogorov41} and refined \citep{kolmogorov62}, the diffusion coefficient has to take the  form \citep{obukhov59}
\begin{equation}
    b_{ij}(\mathbf{x}^\star,t) = \sqrt{C_0 \epsilon(\mathbf{x}^\star,t)} ~ \delta_{ij}~,
    \label{eq:diff}
\end{equation}
where $C_0$ is a dimensionless constant and $\epsilon$ is the local dissipation rate of turbulent kinetic energy into heat. This is especially verified in the high Reynolds number limit (which is relevant in the stellar context), where $C_0$ then actually corresponds to the Kolmogorov constant. This constant is not universal per se; however, it tends asymptotically to a universal value for very high Reynolds numbers, in which case its value is fairly well constrained. An accepted experimental value is $C_0 = 2.1$ \citep{haworthP85}.

For the drift term we adopt the general expression given by the generalised Langevin model \citep{pope83b}
\begin{equation}
    a_i(\mathbf{x}^\star, \mathbf{u}^\star, t) = -\dfrac{1}{\overline{\rho}}\dfrac{\partial \overline{p}}{\partial x_i} + \overline{g_i} + G_{ij}\left(u_j^\star - \widetilde{u_j}\right)~,
    \label{eq:drift}
\end{equation}
where $\overline{\rho}$, $\overline{p}$, and  $\overline{\mathbf{g}}$ are the Reynolds average of the fluid density, gas pressure, and gravitational acceleration, respectively; $G_{ij}$ is a second-order tensor that has the dimension of an inverse time,  to which we   refer as the drift tensor; and $\widetilde{\mathbf{u}}$ is the Favre average of the fluid velocity, with the mass-average (or Favre average) of any quantity $\phi$ being defined as
\begin{equation}
    \widetilde{\phi} \equiv \dfrac{\overline{\rho \phi}}{\overline{\rho}}~.
\end{equation}
All these Reynolds or Favre averages are local and instantaneous quantities, and therefore depend  on both time and space. In Eq. \eqref{eq:drift} they are evaluated at time $t$ and at the position $\mathbf{x}^\star$ of the particle.

The various terms in Eq. \eqref{eq:drift} can be interpreted in the following way. The first two terms are the mean pressure gradient and the gravitational force exerted on the particle, and correspond to the mean force in the momentum equation, the only ones that remain in the absence of turbulence; we note that rotation and magnetic fields are not accounted for in this model. On the other hand, the last term ensures that, were the turbulent sources to disappear, the particle velocity would decay towards the local mean velocity, thus ensuring that the Reynolds stresses are dissipated. More precisely, the drift tensor can be thought of as the rate at which the various Reynolds stresses decay towards zero. In this paper we need not specify the form of the drift tensor, only to say that in the standard approach it is written as a function of the Reynolds stresses, the mean velocity gradients, and the turbulent dissipation \citep{haworthP85}
\begin{equation}
    G_{ij} = f\left(\widetilde{u_i''u_j''}, \partial_i \widetilde{u_j}, \epsilon \right)~,
    \label{eq:Gij_func}
\end{equation}
where $\mathbf{u}'' \equiv \mathbf{u} - \widetilde{\mathbf{u}}$ denotes the fluctuation of the turbulent velocity around its local Favre average. In particular, $G_{ij}$ only depends on the mean fields and not on the particle properties themselves.

Putting together Eqs. \eqref{eq:stochx}, \eqref{eq:stochu}, \eqref{eq:diff}, and \eqref{eq:drift}, we obtain
\begin{align}
    \d x_i^\star &= u_i^\star \d t~, \label{eq:GLMx} \\
    \d u_i^\star &= \left[-\dfrac{1}{\overline{\rho}} \dfrac{\partial \overline{p}}{\partial x_i} + g_i + G_{ij}\left(u_j^\star - \widetilde{u_j}\right)\right] \d t + \sqrt{C_0 \epsilon} \d W_i~. \label{eq:GLMu}
\end{align}
The mean fields $\overline{\rho}$, $\overline{p}$, $\widetilde{\mathbf{u}}$, $G_{ij}$, and $\epsilon$ still need to be closed; we return to this matter in Section \ref{sec:SPH}.

The stochastic equations \eqref{eq:GLMx} and \eqref{eq:GLMu} contain more information than the corresponding average equations on the mean velocity and Reynolds stress tensor, the same way the PDF of a distribution carries more statistical information than its first few moments. We do not make use of these corresponding mean equations in the following; nevertheless, we provide them explicitly in Appendix \ref{app:reynolds-stress}, to which the reader can refer for a better grasp on the origin of the SDE used in this study.

\subsection{From Lagrangian to Eulerian variables\label{sec:lag-to-eul}}

\subsubsection{The Lagrangian mean trajectory formalism}

By construction, the turbulence model given by Eqs. \eqref{eq:GLMx} and \eqref{eq:GLMu} is a Lagrangian model as it pertains to the properties of fluid particles followed along their trajectories. By contrast, we would like to obtain equations on stochastic variables pertaining to the stochastic properties of the flow at a fixed point. This would allow us to ultimately obtain a wave equation where the wave variables can be easily related to the known properties of the modes, something for which a purely Lagrangian\footnote{This statement may seem odd, as Lagrangian variables are actually often used in the analysis of stellar oscillations. However, in this study the term Lagrangian refers to a frame of reference attached to the total velocity of the flow (including both the turbulent velocity and the oscillation velocity), while the usual sense is rather meant to describe a frame of reference attached to the oscillations alone, and actually only ever refers to a pseudo-Lagrangian frame.} description is extremely impractical.

A very general approach to this transcription from Lagrangian to Eulerian variables is the Lagrangian mean trajectories formalism \citep{soward72,andrewsMI78}. In the following, we give the general ideas and the main steps of the derivation; more detailed calculations are provided in Appendix \ref{app:identity-lagrangian}, to which we will refer each time an important step is reached. Let us consider a fluid particle whose time-independent average position is denoted by  $\mathbf{x}$. Its instantaneous position at time $t$ is written as an explicit function of $\mathbf{x}$ and $t$
\begin{equation}
    \mathbf{X}(\mathbf{x},t) = \mathbf{x} + \bm{\xi}(\mathbf{x},t)~,
\end{equation}
where $\bm{\xi}$ is the particle displacement around its mean position\footnote{The variable $\bm{\xi}$ contains the particle displacement due   to the oscillations and to the turbulence. As such, it must not be confused with the fluid displacement due to the oscillations only, and to which the notation $\bm{\xi}$ usually refers.}, the mean position being interpreted as an Eulerian variable.

For any given Eulerian quantity $\phi$, we define its Lagrangian counterpart as
\begin{equation}
    \phi_L(\mathbf{x},t) \equiv \phi(\mathbf{X}(\mathbf{x},t), t)~.
\end{equation}
In particular, we  denote by $\mathbf{u}_L$ the velocity field evaluated at $\mathbf{X}$, in other words the Lagrangian velocity, and by $u_{i,L}$ the $i$-th component of this velocity. Similarly, for any Eulerian averaging process $\langle . \rangle$, we define the corresponding Lagrangian mean $\langle . \rangle_L$ as
\begin{equation}
    \langle \phi \rangle_L \equiv \langle \phi(\mathbf{X}(\mathbf{x},t), t) \rangle~.
\end{equation}
For the time being, we do not yet specify the averaging process $\langle . \rangle$ as this formalism is very general and can be used regardless of how the means are defined. It is important to note here that the mean position $\mathbf{x}$ of the particle is defined in terms of this yet-to-be-determined averaging process. In the following we  simply refer to $\langle . \rangle$ as the `Eulerian mean', but let us keep in mind that it does not necessarily correspond to an ensemble average.

With the above notations and definitions, the following identity can be derived \citep{andrewsMI78}
\begin{equation}
    \left(\dfrac{D\phi}{Dt}\right)_L = \langle D \rangle_L \left(\phi_L \right)~,
    \label{eq:identity}
\end{equation}
where $D/Dt \equiv \partial_t + \mathbf{u} \cdot \bm{\nabla}$ denotes the particle time derivative, and the operator $\langle D \rangle_L$ is defined by
\begin{equation}
    \langle D \rangle_L \equiv \partial_t + \langle \mathbf{u} \rangle_L \cdot \bm{\nabla}~,
\end{equation}
and $\langle \mathbf{u} \rangle_L$ is the Lagrangian mean of the flow velocity. For a detailed derivation of this identity, we refer the reader to Appendix \ref{app:identity-part1}. Because the Lagrangian and Eulerian frames are in motion with respect to one another, the index $_L$ does not commute with either the  space or time derivative. For instance, $\partial (\phi_L) / \partial t$ corresponds to the time derivative of the quantity $\phi$ as seen from the point of view of a fluid parcel (i.e. in the Lagrangian frame), while $(\partial \phi / \partial t)_L$ is the time derivative of the quantity $\phi$ as seen from an Eulerian point of view, and then evaluated at a given Lagrangian coordinate, after the fact. Essentially, Eq. \eqref{eq:identity} describes how the material time derivative commutes with the passage from Lagrangian to Eulerian variables, and will therefore be useful for transcribing our Lagrangian model into a Eulerian one.

Applying Eq. \eqref{eq:identity} on position and velocity respectively yields
\begin{align}
    & \dfrac{\partial \xi_i}{\partial t} = u_{i,L} - \langle u_i \rangle_L - \langle u_j \rangle_L \partial_j\xi_i~, \label{eq:GLM_eulx} \\
    & \dfrac{\partial \left(u_{i,L}\right)}{\partial t} = \left(\dfrac{\partial u_i}{\partial t}\right)_L + \left[u_{j,L}\delta_{jk} - \langle u_j \rangle_L \delta_{jk} - \langle u_j \rangle_L \partial_j \xi_k \right] \left(\dfrac{\partial u_i}{\partial x_k}\right)_L~. \label{eq:lag-to-eulu}
\end{align}
The derivation of these two equations is given in detail in Appendix \ref{app:identity-part2}. We note that, for the moment, the displacement $\bm{\xi}$ and velocity $\mathbf{u}$ are flow variables, which is why they are not denoted with a $^\star$. We now relate these flow quantities to the position $\mathbf{x}^\star$ and velocity $\mathbf{u}^\star$ of the fluid particles. Since $\bm{\xi}$ and $\mathbf{u}_L$ correspond to the displacement and velocity of the particle whose mean position is $\mathbf{x}$, then for any fixed $\mathbf{x}$ we have
\begin{align}
    \mathbf{x}^\star(t) &= \mathbf{x} + \bm{\xi}(\mathbf{x},t)~, \label{eq:change-varx} \\
    \mathbf{u}^\star(t) &= \mathbf{u}_L(\mathbf{x},t)~, \label{eq:change-varu}
\end{align}
so that
\begin{align}
    & \dfrac{\d \mathbf{x}^\star}{\d t} = \dfrac{\partial \bm{\xi}}{\partial t}~, \label{eq:star-to-eulx} \\
    & \dfrac{\d \mathbf{u}^\star}{\d t} = \dfrac{\partial \left(\mathbf{u}_L\right)}{\partial t}~. \label{eq:star-to-eulu}
\end{align}
Putting together Eqs. \eqref{eq:GLMu}, \eqref{eq:lag-to-eulu}, and \eqref{eq:star-to-eulu} we obtain
\begin{multline}
    \left(\dfrac{\partial u_i}{\partial t}\right)_L + \left[ u_{j,L} - \langle u_j \rangle_L - \langle u_k \rangle_L \left(\dfrac{\partial \xi_j}{\partial x_k}\right)_L\right] \left(\dfrac{\partial u_i}{\partial x_j}\right)_L \\
    = -\dfrac{1}{\overline{\rho}_L} \left(\dfrac{\partial\overline{p}}{\partial x_i}\right)_L + g_{i,L} + G_{ij,L} \left(u_{j,L} - \widetilde{u_j}_{,L}\right) + \sqrt{C_0 \epsilon_L} \eta_{i,L}~. \label{eq:GLM_eulu}
\end{multline}
By construction, $\bm{\eta}(\mathbf{x},t) \equiv \d \mathbf{W} / \d t$ is a multi-variate Gaussian process whose values at two distinct locations are completely uncorrelated, and which verifies\footnote{The stochastic process $\mathbf{W}(t)$ is not defined as an ordinary function of time, and therefore its derivative $\bm{\eta}$ cannot be defined the classical way; in fact, $\mathbf{W}$ is nowhere differentiable, as can be seen from Eq. \eqref{eq:W-var}. However, $\bm{\eta}$ can be defined formally, with its statistical properties given in the sense of distributions rather than ordinary functions. These definitions are at the heart of Ito stochastic calculus \citep[][Chap 4]{gardiner85book}.}
\begin{align}
    & \overline{\bm{\eta}(\mathbf{x},t)} = \mathbf{0}~, \label{eq:eta-mean} \\
    & \overline{\eta_i(\mathbf{x},t') \eta_j(\mathbf{x},t)} = \delta(t' - t)\delta_{ij}~, \label{eq:eta-var}
\end{align}
where $\delta(t)$ is the Dirac distribution.

Finally, we note that in this form, all the quantities present in Eq. \eqref{eq:GLM_eulu} are evaluated at the instantaneous position $\mathbf{X}$. Insofar as the transformation $\mathbf{x} \mapsto \mathbf{X}$ is invertible (i.e. for any $\mathbf{x}$ and $t$ there exists $\mathbf{y}$ such that $\mathbf{X}(\mathbf{y},t) = \mathbf{x}$), we can drop the notation $_L$ from this equation,  except in $\langle \mathbf{u} \rangle_L$, thus yielding a stochastic equation for the evolution of the Eulerian velocity $\mathbf{u}$.

\subsubsection{Specification of the averaging process}

For the moment, we still have not specified the nature of the mean which defines both the mean particle position $\mathbf{x}$ and the Lagrangian mean velocity $\langle \mathbf{u} \rangle_L$. The specification of this averaging process is crucial, and the fact that this formalism applies to any averaging process is of the utmost importance. Usually, the Lagrangian mean trajectory formalism is used to transform the exact Eulerian hydrodynamics equations into equations on carefully defined Lagrangian means, which happen to be much more suitable to the study of hydrodynamic waves \citep{andrewsMI78}. In that context it is customary to consider that $\langle . \rangle$ actually denotes an ensemble average, so that the mean values contain the oscillations as well as the background equilibrium, whereas the fluctuations contain the turbulent fields.

In the present context, however, this is not the picture after which we are. Instead, {we want the fluctuating part to contain the waves in addition to the turbulent fluctuations, whereas the means should only contain the background equilibrium}. Only then can we obtain a wave equation directly containing turbulence-induced fluctuations, and therefore the turbulence--oscillation coupling. As such, we will define $\langle . \rangle$ as a time average over timescales that are very long compared to the typical turbulent timescale, the period of the oscillations, and their lifetime. This ensures that the mean values only contain the time-independent equilibrium, and the fluctuating part does indeed contain both the waves and the turbulent fields.

\begin{figure}
    \centering
    \includegraphics[width=\linewidth]{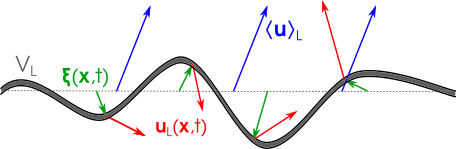}
    \caption{Visualisation of the fluid displacement $\bm{\xi}$ (in green), Lagrangian velocity $\mathbf{u}_L$ (in red), and Lagrangian mean velocity $\langle \mathbf{u} \rangle_L$ (in blue), in the case where the mean $\langle . \rangle$ is defined as an average over a given axis (spanning horizontally in the figure). The grey volume $V_L$ represents a thin tube of fluid initially lying along the horizontal direction. The fluid displacement and velocity $\bm{\xi}$ and $\mathbf{u}_L$ pertain to the deformation and local velocity of the tube, while $\langle \mathbf{u} \rangle_L$ refers to the velocity of the centre  of mass of the tube, represented by the dashed horizontal line. This illustration is inspired by \citet{andrewsMI78} (see bottom panel of their Figure 1).}
    \label{fig:dessin}
\end{figure}

The fact that $\langle . \rangle$ denotes a time average also considerably simplifies Eqs. \eqref{eq:GLM_eulx} and \eqref{eq:GLM_eulu}. The Lagrangian mean velocity $\langle \mathbf{u} \rangle_L$ is constructed in such a way that when the fluid velocity at $\mathbf{X}$ is $\mathbf{u}_L$, then the mean position $\mathbf{x}$ is displaced with the velocity $\langle \mathbf{u} \rangle_L$ \citep{andrewsMI78}. A perhaps more illustrative way of interpreting the quantities $\bm{\xi}$, $\mathbf{u}_L$, and $\langle \mathbf{u} \rangle_L$ is given in Fig. \eqref{fig:dessin}, in the case where $\langle . \rangle$ is defined as a spatial average in a given direction $x$. If we isolate a thin tube of fluid lying along this axis, then $\bm{\xi}$ corresponds to the local deformation of the tube, $\mathbf{u}_L$ corresponds to the instantaneous velocity of the local portion of tube, and $\langle \mathbf{u} \rangle_L$ corresponds to the velocity of the centre of mass of the tube.

In our case where `average' refers to time average, $\langle \mathbf{u} \rangle_L$ refers to the movement of the centre of mass of a given parcel of fluid in the absence of turbulent convection and oscillations, or in other words, to the background fluid velocity. We note that the effects of rotation, if we were to take them into account, would be encompassed in $\langle \mathbf{u} \rangle_L$. In the absence of rotation, however, we have $\langle \mathbf{u} \rangle_L = \mathbf{0}$, and Eqs. \eqref{eq:GLM_eulx} and \eqref{eq:GLM_eulu} then reduce to
\begin{align}
    & \dfrac{\partial \xi_i}{\partial t} = u_i(\mathbf{x}+\mathbf{\bm{\xi}},t)~, \label{eq:GLM_eulx_v0} \\
    & \dfrac{\partial u_i}{\partial t} + u_j \dfrac{\partial u_i}{\partial x_j} = -\dfrac{1}{\overline{\rho}} \dfrac{\partial\overline{p}}{\partial x_i} + g_i + G_{ij} \left(u_j - \widetilde{u_j}\right) + \sqrt{C_0 \epsilon} ~ \eta_i~. \label{eq:GLM_eulu_v0}
\end{align}
It must be noted that while the velocity appearing on the right-hand side of Eq. \eqref{eq:GLM_eulx_v0} is evaluated at the Lagrangian position $\mathbf{x} + \bm{\xi}(\mathbf{x},t)$, every quantity in Eq. \eqref{eq:GLM_eulu_v0}, by contrast, is evaluated at the Eulerian position $\mathbf{x}$. Furthermore, it should be noted that everything depends on time, even when not explicitly specified. The key difference between Eqs. \eqref{eq:GLMx} and \eqref{eq:GLMu}, on the one hand, and \eqref{eq:GLM_eulx_v0} and \eqref{eq:GLM_eulu_v0} on the other, is that the variables whose evolution is described are no longer the Lagrangian quantities $\mathbf{x}^\star(t)$ and $\mathbf{u}^\star(t)$ pertaining to a set of fluid particles, but the Eulerian quantities $\bm{\xi}(\mathbf{x},t)$ and $\mathbf{u}(\mathbf{x},t)$ pertaining to a set of Eulerian fixed positions $\mathbf{x}$. This description will allow for a much more practical derivation of the wave equation in Section \ref{sec:lin}.

In particular, the momentum equation now makes the contribution of the turbulent pressure explicitly appear, in the form of the advection term on the left-hand side of Eq. \eqref{eq:GLM_eulu_v0}. The contribution of the turbulent fluctuations of the gas pressure and the turbulent dissipation, on the other hand, are still contained in the last two terms on the right-hand side. Furthermore, the procedure described in this section allowed us to rigorously separate the effect of the equilibrium background (defined by a time average over very long timescales) from the joint contributions of the turbulence and of the oscillations (both of which are contained in the fluctuations around the equilibrium background), and thus will allow us to study their mutual coupling in a consistent framework.

\subsection{Evaluating the mean fields: Smoothed particle hydrodynamics\label{sec:SPH}}

The stochastic model on fluid displacement $\bm{\xi}$ and velocity $\mathbf{u}$, comprised of Eqs. \eqref{eq:GLM_eulx_v0} and \eqref{eq:GLM_eulu_v0}, is still not in closed form, as it contains several mean quantities (mean density $\overline{\rho}$, gas pressure $\overline{p}$, and velocity $\widetilde{\mathbf{u}}$), as well as the turbulent dissipation $\epsilon$ appearing in the diffusion tensor, and the Reynolds stress tensor $\widetilde{u_i''u_j''}$ and shear tensor $\partial_i\widetilde{u_j}$ on which the drift tensor $G_{ij}$ depends. These equations must therefore be supplemented with a model for the mean fields.

One possibility is to make use of a large-eddy simulation (LES) (or direct numerical simulation) to simulate the large-scale flow using the exact equations of hydrodynamics, and then use the mean fields yielded by this simulation as external inputs into the stochastic model. However, the mean fields appearing in Eq. \eqref{eq:GLM_eulu_v0} are instantaneous averages (for example, $\overline{\rho}$ is the Reynolds-averaged density at a given time $t$), not time averages. As a result, the ergodic principle cannot be used to extract the means from a LES. The only way to do this would be to consider that horizontal averages in a 3D LES provide   an accurate estimate of instantaneous ensemble averages, and would only work in the scope of a 1D model. Furthermore, this procedure would defeat the purpose of what we are trying to achieve;  since the mean fields contain the information on the oscillations, without containing the turbulence, treating them as external inputs would effectively amount to modelling the turbulence and the oscillations in a separate manner, which is what we are trying to avoid.

An alternative method makes use of the particle representation we adopt in Section \ref{sec:GLM}. The set of fluid particles used to represent the flow contains all the required statistical information, so that the mean fields can actually be estimated directly from the set of fluid particles themselves \citep{weltonP97}. This is the core idea behind particle methods, and particularly smoothed particle hydrodynamics (SPH). The reader can refer to \citet{liuL10review} or \citet{monaghan92review} for a comprehensive review on the subject, or to \citet{springel10review} for the use of SPH in the astrophysical context, but we give an outline of this method in the following.

Ideally, we would like to estimate all local means at a given Eulerian position $\mathbf{x}$ by averaging the corresponding particle-level quantity over all fluid particles {conditioned on their being located at $\mathbf{x}$}. However, implementing this last condition exactly does not yield the required result:  for any given position $\mathbf{x}$, any individual fluid particle has exactly zero probability of finding itself at this exact location. Therefore, it is necessary to relax the condition on particle position, and instead of computing means over particles exactly located at $\mathbf{x}$, we compute them over particles within a given compact-support vicinity of $\mathbf{x}$.

Thus, a kernel function $K(\mathbf{r})$ is introduced, which serves as a weighting function to implement the particle-position condition in the estimation of the means. The exact form of $K$ is not important here, but we  mention some of its properties, namely that  it is a compact-support function, it is normalised to unity, and it is isotropic. The first two properties are mandatory, as the first one ensures that distant particles cannot impact local means, and the second that the estimation of the means is unbiased. The third property makes the subsequent calculations much easier to carry out. A good example is the kernel function used by \citet{weltonP97}
\begin{equation}
    \begin{array}{lll}
        K(\mathbf{r}) &= c\left(1 + 3\dfrac{|\mathbf{r}|}{h}\right)\left(1 - \dfrac{|\mathbf{r}|}{h}\right)^3 ~ & ~ \text{if} ~ |\mathbf{r}| < h~, \\
        &= 0 ~ & ~ \text{if} ~ |\mathbf{r}| > h~,
    \end{array}
\end{equation}
where $\mathbf{r}$ is the position of the particle with respect to the centre of the kernel (where the mean is estimated), $c = 105 / (16\pi h^3)$ is defined by the normalisation condition\footnote{The reason the value of $c$ given here is different from the value $c=4/(5h)$ given in \citet{weltonP97} is that they considered the 1D case, whereas we consider the 3D case.}, and $h$ is the size of the kernel compact support. This expression ensures that the kernel function and its first two derivatives are continuous at the surface of its support.

The SPH formalism is best formulated if we temporarily return to the representation of the flow as a large set of $N$ particles, whose position and velocity we denote by $\mathbf{x}^{\star(i)}$ and $\mathbf{u}^{\star(i)}$, respectively, where $i$ is the index used to identify each particle. For any quantity $Q$ pertaining to the flow representation, if there is an equivalent quantity $Q^\star$ in the particle representation, we can estimate the mean value of $Q$ at any Eulerian position $\mathbf{x}$ and time $t$ through the following kernel estimator
\begin{equation}
    \overline{Q(\mathbf{x},t)} = \bigsum{i} \dfrac{\Delta m^{(i)}}{\rho^{(i)}} Q^{\star(i)}(t) K\left(\mathbf{x}^{\star(i)}(t) - \mathbf{x}\right)~,
    \label{eq:SPH}
\end{equation}
where $\Delta m^{(i)}$ is the mass carried by the particle $i$, and $\rho^{(i)}$ is the mass density characterising particle $i$. As such, the quantity $\Delta m^{(i)} / \rho^{(i)}$ appearing under the sum corresponds to the lumped volume of fluid that the particle represents.

Setting $Q^\star = \rho$, $Q^\star = \rho \mathbf{u}^\star$, and $Q^\star = \rho (u_i^\star - \widetilde{u_i})(u_j^\star - \widetilde{u_j})$ alternatively in Eq. \eqref{eq:SPH}, we find respectively the Reynolds-averaged density, the mass-averaged velocity, and the Reynolds stress tensor
\begin{align}
    & \overline{\rho}(\mathbf{x},t) = \bigsum{i} \Delta m^{(i)} K(\mathbf{x}^{\star(i)}(t) - \mathbf{x})~, \label{eq:SPH_disc_rho} \\
    \vspace{0.1cm} \nonumber \\
    & \widetilde{\mathbf{u}}(\mathbf{x},t) = \dfrac{1}{\overline{\rho}(\mathbf{x},t)} \bigsum{i} \Delta m^{(i)} \mathbf{u}^{\star(i)}(t) K(\mathbf{x}^{\star(i)}(t) - \mathbf{x})~, \label{eq:SPH_disc_u} \\
    \vspace{0.1cm} \nonumber \\
    & \widetilde{u_i'' u_j''}(\mathbf{x},t) = \dfrac{1}{\overline{\rho}(\mathbf{x},t)} \bigsum{i} \Delta m^{(i)} \left(u_i^{\star(i)}(t) - \widetilde{u_i}(\mathbf{x},t)\right) \nonumber \\
    & \hspace{2.5cm} \times \left(u_j^{\star(i)}(t) - \widetilde{u_j}(\mathbf{x},t)\right) K(\mathbf{x}^{\star(i)}(t) - \mathbf{x})~. \label{eq:SPH_disc_rey}
\end{align}
In particular, we note that in the SPH formalism, the local mean density is computed by counting the particles present in the vicinity. This means that the continuity condition is automatically met in the particle representation, thus lowering the order of the set of equations needed to describe the flow.

We then rewrite Eqs. \eqref{eq:SPH_disc_rho}, \eqref{eq:SPH_disc_u}, and \eqref{eq:SPH_disc_rey} in the representation chosen in Section \ref{sec:lag-to-eul}, specifically in terms of $\bm{\xi}(\mathbf{x})$ and $\mathbf{u}(\mathbf{x})$ rather than $\mathbf{x}^{\star(i)}$ and $\mathbf{u}^{\star(i)}$. To do so, we perform the change of variables given by Eqs. \eqref{eq:change-varx} and \eqref{eq:change-varu}. Furthermore, the sum over infinitesimally small masses can be replaced by a continuous integral over $\d m \equiv \rho_0(\mathbf{y}) \d^3\mathbf{y}$, where $\rho_0$ is the equilibrium fluid density (which can be thought of as an average of the local fluid density over very long timescales so as to only contain the background value). Finally, in this new representation, the SPH formalism yields\begin{align}
    & \overline{\rho}(\mathbf{x},t) = \displaystyle\int \d^3\mathbf{y} ~ \rho_0(\mathbf{y}) K(\mathbf{y} + \bm{\xi}(\mathbf{y},t) - \mathbf{x})~, \label{eq:SPH_cont_rho} \\
    \vspace{0.1cm} \nonumber \\
    & \widetilde{\mathbf{u}}(\mathbf{x},t) = \dfrac{1}{\overline{\rho}(\mathbf{x},t)} \displaystyle\int \d^3\mathbf{y} ~ \rho_0(\mathbf{y}) \mathbf{u}(\mathbf{y} + \bm{\xi}(\mathbf{y},t),t) K(\mathbf{y} + \bm{\xi}(\mathbf{y},t) - \mathbf{x})~, \label{eq:SPH_cont_u} \\
    \vspace{0.1cm} \nonumber \\
    & \widetilde{u_i'' u_j''}(\mathbf{x},t) = \dfrac{1}{\overline{\rho}(\mathbf{x},t)} \displaystyle\int \d^3\mathbf{y} ~ \rho_0(\mathbf{y}) \left(\vphantom{u_j} u_i(\mathbf{y} + \bm{\xi}(\mathbf{y},t),t) - \widetilde{u_i}(\mathbf{y},t)\right) \nonumber \\
    & \hspace{1.0cm} \times \left(u_j(\mathbf{y} + \bm{\xi}(\mathbf{y},t),t) - \widetilde{u_j}(\mathbf{y},t)\right) K(\mathbf{y} + \bm{\xi}(\mathbf{y},t) - \mathbf{x})~. \label{eq:SPH_cont_rey}
\end{align}
While these integrals span across the entire volume of the star, we note that they actually only involve the compact support vicinity of $\mathbf{x}$ defined by the kernel function $K$.

Two more points need to be addressed here. The first   concerns the mean pressure $\overline{p}$. For the sake of simplicity, we consider a polytropic relation between the gas pressure and density in the form
\begin{equation}
    \ln\left(\dfrac{\overline{p}(\mathbf{x},t)}{p_0(\mathbf{x})}\right) =  \gamma(\mathbf{x}) \ln\left(\dfrac{\overline{\rho}(\mathbf{x},t)}{\rho_0(\mathbf{x})}\right)~,
    \label{eq:SPH_p}
\end{equation}
where $p_0$ is the equilibrium gas pressure (defined in the same way as $\rho_0$), and we allow the polytropic exponent $\gamma$ to depend on space. We note that we do not consider the possibility that the oscillations may entail fluctuations in the polytropic index itself. We also note that we can recover the isentropic case at any point by setting $\gamma = \Gamma_1$, where $\Gamma_1$ is the equilibrium first adiabatic exponent.

The second point concerns the turbulent dissipation rate $\epsilon$, or equivalently the turbulent frequency $\omega_t$ defined through
\begin{equation}
    \omega_t \equiv \dfrac{\epsilon}{k} = \dfrac{2\epsilon}{\widetilde{u_i''u_i''}}~,
    \label{eq:SPH_epsilon}
\end{equation}
where $k$ is the turbulent kinetic energy. Physically, $\omega_t$ can be interpreted as the inverse of the characteristic lifetime associated with the energy-bearing eddies. The turbulent kinetic energy $k$ is given in closed form by the velocity part of the model (here it is given by half the trace of Eq. \ref{eq:SPH_cont_rey}), and we still need to model $\omega_t$. Usually, this is done either by adding a model equation for the mass-averaged dissipation rate $\widetilde{\omega_t}$, which is very similar to the approach followed in two-equation models of turbulence, such as the $k-\epsilon$ model \citep{jonesL72}, or else by adding $\omega_t^\star$ to the particle properties in the Lagrangian stochastic model, such as in the refined Langevin model \citep{popeC90}. However, in the present work, and in the scope of the generalised Langevin model, we  regard $\omega_t$ as a time-independent equilibrium quantity, which can still, however, depend on $\mathbf{x}$. Physically, this amounts to assuming that all eddies have  the same typical lifetime, regardless of their size, but that it can depend on the depth at which they are located. In the long run, it will be necessary to go beyond this drastic assumption.

To recap Section \ref{sec:stoch-model}, the model equations are Eqs. \eqref{eq:GLM_eulx_v0} and \eqref{eq:GLM_eulu_v0}, which are stochastic differential equations governing the evolution of the fluid displacement $\bm{\xi}(\mathbf{x},t)$ and the Eulerian velocity $\mathbf{u}(\mathbf{x},t)$ for any given Eulerian position $\mathbf{x}$. The mean density $\overline{\rho}$, the mass-averaged velocity $\widetilde{\mathbf{u}}$, and the Reynolds stress tensor $\widetilde{u_i''u_j''}$ are given by Eqs. \eqref{eq:SPH_cont_rho}, \eqref{eq:SPH_cont_u}, and \eqref{eq:SPH_cont_rey}, respectively, as explicit functions of $\bm{\xi}(\mathbf{x},t)$ and $\mathbf{u}(\mathbf{x},t)$ only. The mean pressure is given by Eq. \eqref{eq:SPH_p} as a function of mean density, and the turbulent dissipation rate $\epsilon$ is given by Eq. \eqref{eq:SPH_epsilon}. Therefore, all the quantities appearing in the model equations are written explicitly as functions of the modelled variables $\bm{\xi}(\mathbf{x},t)$ and $\mathbf{u}(\mathbf{x},t)$ themselves: the model is in closed form. The only inputs of the model are 1) the equilibrium density $\rho_0(\mathbf{x})$, gas pressure $p_0(\mathbf{x})$, and polytropic exponent $\gamma(\mathbf{x})$, which can be extracted from an equilibrium model of the star; 2) the functional form of the drift tensor $G_{ij}$ (see Eq. \ref{eq:Gij_func}), which can be constrained using direct numerical simulations or experimental measurements \citep{pope94}; and 3) the equilibrium turbulent frequency $\omega_t(\mathbf{x})$, which can be constrained using a 3D hydrodynamic simulation of the atmosphere of the star, or on the contrary serve as a control parameter for turbulence, which can be varied for a parametric study.

\section{The stochastic wave equation\label{sec:waveeq}}

We now set out to linearise the closed set of equations derived in Section \ref{sec:stoch-model} to obtain a linear stochastic wave equation that was  designed to govern the physics of the mode while simultaneously encompassing the effect of turbulence on the mode. We then discuss the properties of this wave equation, and how it relates to other forms of the wave equation obtained in previous studies.

\subsection{Linearisation of the stochastic model\label{sec:lin}}

The system to linearise is comprised of Eqs. \eqref{eq:GLM_eulx_v0}, \eqref{eq:GLM_eulu_v0}, \eqref{eq:SPH_cont_rho}, \eqref{eq:SPH_cont_u}, \eqref{eq:SPH_cont_rey}, \eqref{eq:SPH_p}, and \eqref{eq:SPH_epsilon}. The only variables in these equations are the fluid displacement $\bm{\xi}(\mathbf{x},t)$ and the Eulerian velocity $\mathbf{u}(\mathbf{x},t)$. We define
\begin{align}
    & \xiosc(\mathbf{x},t) \equiv \bm{\xi}(\mathbf{x},t) - \bm{\xi}_t(\mathbf{x},t)~, \label{eq:split-varx} \\
    & \uosc(\mathbf{x},t) \equiv \mathbf{u}(\mathbf{x},t) - \ut(\mathbf{x},t)~, \label{eq:split-varu}
\end{align}
where $\bm{\xi}_t$ and $\ut$ are the fluid displacement and velocity that would be obtained if there were no oscillations, and that represent the turbulent component of the fluid displacement and velocity, and $\xiosc$ and $\uosc$ are the oscillatory displacement and velocity. We note  that {while the variables are split into a turbulent and an oscillatory component, the system of equations itself is not split into equations for turbulence and equations for oscillations}; instead, there is still only  one system of equations containing both components. This is in contrast, for instance, with MLT or Reynolds-Averaged Navier-Stokes (RANS) approaches where the equations are averaged from the start, thus implicitly separating the two components whose coupling we wish to study.

Obtaining a linear wave equation requires the adoption of a certain number of hypotheses regarding the fluid variables, which we itemise here.

\textbf{(H1)} We consider $|\uosc| \ll |\ut|$. This ordering is justified by the fact that, at the top of the convective envelope of solar-like oscillators, the typical turbulent velocities have much higher amplitudes than the oscillatory velocities;   the former are of the order of a few km.s$^{-1}$, while the latter are of the order of a few tens of cm.s$^{-1}$. This allows us to treat $\uosc$ as a first-order perturbation compared to $\ut$, and any second- or higher-order occurrence of $\uosc$ will be discarded.

\textbf{(H2)} We consider $|\xiosc| \ll h,H_p~~$, where we recall that $h$ is the size of the averaging kernel function $K$, and $H_p \equiv -(\d\ln p_0 / \d r)^{-1}$ is the pressure scale height. In other words, the modal fluid displacement is much smaller than the stratification length scale, and the width of the kernel function must be sufficiently large. The first hypothesis is justified by the fact that, in the Sun for instance, the modal displacement is of the order of a few tens of meters, while $H_p$ is of the order of a few hundreds of kilometers. The second hypothesis, on the other hand, constitutes a constraint on $h$. This allows us to treat $\xiosc$ as a first-order perturbation compared to all length scales relevant to the problem, and any second- or higher-order occurrence of $\xiosc$ will be discarded.

\textbf{(H3)} We  adopt the anelastic approximation for turbulence, in the sense that we  consider $\rho_t \ll \rho_0$, where $\rho_t$ is the turbulent fluctuation of density, and $\rho_0$ the equilibrium density. This is the most severe approximation we make in this section. Nevertheless, the anelastic approximation is widely used in analytical models of turbulent convection in these regions, on the grounds that the flow is subsonic (with turbulent Mach numbers peaking at around $0.3$ in the superadiabatic region), as shown by 3D hydrodynamic simulations of the atmosphere of these stars \citep{nordlund09review}. Using the continuity equation, this amounts to neglecting the quantity $\bm{\nabla} \cdot \left(\rho_0 \bm{\xi}_t\right)$. As will become apparent in the following, this   allows us to discard all $\bm{\xi}_t$-dependent contributions in the linearisation of the ensemble averages in the SPH formalism.

\textbf{(H4)} We   consider that the turbulent velocity field $\ut$ is the same as   it would be without the presence of an oscillating velocity $\uosc$;  in other words, we neglect the back-reaction of the oscillations on the turbulent motions of the gas. We justify this approximation in Appendix \ref{app:backreaction} on the basis of a discussion that can be found in \citet[][see their Section 5.1.1]{buhler09book}. We note that the back-reaction being neglected here concerns both the equilibrium part and the stochastic part (i.e. both the equilibrium structure of the star and the turbulent velocity field). This assumption allows us to consider $\ut$ as an input to the model, whose statistical properties (average, covariance, autocorrelation function) are considered completely known.

\textbf{(H5)} We  consider that the gravitational potential is not perturbed   by the turbulent motions of the gas or by its oscillatory motions. These are actually two separate approximations. The first   is justified by the fact that the Reynolds-averaged mass flow through any given horizontal layer due to turbulence is zero, meaning the total mass present beneath this layer is always the same. The second   corresponds to the Cowling approximation, and is justified for modes that feature a large number of radial nodes. These two approximations put together allow us to replace the gravitational acceleration $\mathbf{g}$ by its equilibrium value $\mathbf{g}_0$, which only depends on the hydrostatic equilibrium of the star.

We insist on the fact that these approximations,  with the exception of (H5), only concern the fluid displacement and velocity. By contrast, {no specific approximation is adopted concerning the mean fields;  a linearised form of these mean fields naturally arises from the SPH formalism and the hypotheses (H1) through (H4).} As an example, let us consider the mean density $\overline{\rho}$. Plugging Eq. \eqref{eq:split-varx} into Eq. \eqref{eq:SPH_cont_rho}, we find
\begin{equation}
    \overline{\rho}(\mathbf{x},t) = \displaystyle\int \d^3\mathbf{y} ~ \rho_0(\mathbf{y}) K\left(\mathbf{y} + \bm{\xi}_t(\mathbf{y},t) + \xiosc(\mathbf{y},t) - \mathbf{x}\right)~.
\end{equation}
Then, using (H2) to linearise in terms of the displacement, we find
\begin{multline}
    \overline{\rho}(\mathbf{x},t) = \displaystyle\int \d^3\mathbf{y} ~ \rho_0(\mathbf{y}) K(\mathbf{y} - \mathbf{x}) + \displaystyle\int \d^3\mathbf{y} ~ \rho_0(\mathbf{y}) \left.\left(\xiosc(\mathbf{y},t)\cdot\bm{\nabla}\right) K\right|_{\mathbf{y} - \mathbf{x}} \\
    + \displaystyle\int \d^3\mathbf{y} ~ \rho_0(\mathbf{y}) \left.\left(\bm{\xi}_t(\mathbf{y},t)\cdot\bm{\nabla}\right) K\right|_{\mathbf{y} - \mathbf{x}}~.
\end{multline}
The first term on the right-hand side corresponds to the kernel estimate of $\rho_0$ at $\mathbf{x}$. By construction, kernel estimation is a representation of ensemble averaging, but $\rho_0$ is already an equilibrium quantity, and therefore is equal to its own ensemble average. Furthermore, the last term on the right-hand side can be discarded on account of hypothesis (H3). Performing an integration by part makes the quantity $\bm{\nabla} \cdot \left(\rho_0\bm{\xi}_t \right)$ appear under the integral sign. Therefore, we eventually find
\begin{equation}
    \overline{\rho}(\mathbf{x},t) = \rho_0(\mathbf{x}) + \underbrace{\displaystyle\int \d^3\mathbf{y} ~ \rho_0(\mathbf{y}) \left.\left(\xiosc(\mathbf{y},t)\cdot\bm{\nabla}\right) K\right|_{\mathbf{y} - \mathbf{x}}}_{\equiv \rho_1(\mathbf{x},t)}~.
    \label{eq:example_rho}
\end{equation}
The quantity $\rho_1$ represents the Eulerian modal fluctuations of density, but of important note is the fact that {at no point did we explicitly decompose $\overline{\rho}$ into an equilibrium value $\rho_0$ and a residual, oscillatory part;  instead, the decomposition \eqref{eq:example_rho} arises naturally from the SPH formalism, and hypotheses (H2) and (H3)}.

The linear wave equation is derived in detail in Appendix \ref{app:lin} using the hypotheses listed above. Ultimately, we obtain
\begin{align}
    & \dfrac{\partial \xiosc}{\partial t} - \uosc - (\xiosc \cdot \bm{\nabla})\ut - (\bm{\xi_t} \cdot \bm{\nabla})\uosc = (\bm{\xi_t} \cdot \bm{\nabla})\ut~, \label{eq:xiosc} \\
    & \dfrac{\partial \uosc}{\partial t} - \mathbf{L}_1^d - \mathbf{L}_1^s = \mathbf{L}_0~,
    \label{eq:uosc}
\end{align}
where
\begin{align}
    L_{1,i}^d &= \left[ \dfrac{1}{\rho_0}\dfrac{\partial p_0}{\partial x_i} - \dfrac{\partial c_0^2}{\partial x_i} \right] \dfrac{1}{\rho_0}\displaystyle\int \d^3\mathbf{y} ~ \rho_0(\mathbf{y}) \left.\left(\xiosci{j} \partial_j K^{\mathbf{x}}\right)\right|_{\mathbf{y},t} \nonumber \\
    & + \dfrac{c_0^2}{\rho_0} \displaystyle\int \d^3\mathbf{y} ~ \rho_0(\mathbf{y}) \left.\left(\xiosci{j} \partial_j \partial_i K^{\mathbf{x}}\right)\right|_{\mathbf{y},t} \nonumber \\
    & + G_{ij,0}\left( \uosci{j} - \dfrac{1}{\rho_0(\mathbf{x})}\displaystyle\int \d^3\mathbf{y} ~ \rho_0(\mathbf{y}) \uosci{j}(\mathbf{y}) K^{\mathbf{x}}(\mathbf{y}) \right)~,
    \label{eq:L1d}
\end{align}
\begin{align}
    L_{1,i}^s &= -\uosci{j} \partial_j \uti{i} - \uti{j} \partial_j \uosci{i} \nonumber \\
    & -G_{ij,0} \dfrac{1}{\rho_0}\displaystyle\int \d^3\mathbf{y} ~ \rho_0(\mathbf{y}) \left.\left( \xiosci{k} \partial_k \left(\uti{j} K^{\mathbf{x}} \right) \right)\right|_{\mathbf{y},t} \nonumber \\
    & + \left(\dfrac{\partial G_{ij}}{\partial \widetilde{u_k'' u_l''}} \widetilde{u_k'' u_l''}_1 + \dfrac{\partial G_{ij}}{\partial (\partial_k\widetilde{u_l})} \partial_k \widetilde{u_l} + \dfrac{\partial G_{ij}}{\partial \epsilon} \omega_t k_1 \right) \uti{j} \nonumber \\
    & + \dfrac{1}{2}\sqrt{\dfrac{C_0 \omega_t}{k_0}} k_1 \eta_i~,
    \label{eq:L1s}
\end{align}
\begin{equation}
    L_{0,i} = -\dfrac{1}{\rho_0} \dfrac{\partial \left( \rho_0 \uti{i}\uti{j} - \rho_0 \overline{\uti{i} \uti{j}} \right)}{\partial x_j}~,
    \label{eq:L0}
\end{equation}
\begin{equation}
    \begin{array}{ll}
        \widetilde{u_i'' u_j''}_1 =& - \dfrac{\widetilde{u_i''u_j''}_0}{\rho_0} \displaystyle\int \d^3\mathbf{y} ~ \rho_0 \xiosci{k} \partial_k K^{\mathbf{x}} \\
        \vspace{0.05cm} \\
        & + \dfrac{1}{\rho_0}\displaystyle\int \d^3\mathbf{y} ~ \rho_0 \xiosci{k} \partial_k \left(\uti{i} \uti{j} K^{\mathbf{x}}\right) \\
        \vspace{0.05cm} \\
        & + \dfrac{1}{\rho_0}\displaystyle\int \d^3\mathbf{y} ~ \rho_0 \uti{i} \uosci{j} K^{\mathbf{x}} \\
        \vspace{0.05cm} \\
        & + \dfrac{1}{\rho_0}\displaystyle\int \d^3\mathbf{y} ~ \rho_0 \uti{j} \uosci{i} K^{\mathbf{x}}~,
    \end{array}
    \label{eq:rey1}
\end{equation}
\begin{align}
    (\partial_i \widetilde{u_j})_1 &= - \dfrac{1}{\rho_0} \displaystyle\int \d^3\mathbf{y} ~ \rho_0 \uosci{j} \partial_i K^{\mathbf{x}} - \dfrac{1}{\rho_0} \displaystyle\int \d^3\mathbf{y} ~ \rho_0 \xiosci{k} \partial_k \uti{j} \partial_i K^{\mathbf{x}} \nonumber \\
    & - \dfrac{1}{\rho_0^2} \dfrac{\partial \rho_0}{\partial x_i} \displaystyle\int \d^3\mathbf{y} ~ \rho_0 \uosci{j} K^{\mathbf{x}} \nonumber \\
    & - \dfrac{1}{\rho_0^2} \dfrac{\partial \rho_0}{\partial x_i} \displaystyle\int \d^3\mathbf{y} ~ \rho_0 \xiosci{k} \partial_k \left(\uti{j} K^{\mathbf{x}}\right)~,
    \label{eq:shear1}
\end{align}
\begin{equation}
    k_1 = \dfrac{1}{2}\widetilde{u_i''u_i''}~,
\end{equation}
where $c_0^2 \equiv p_0\gamma / \rho_0$ is the equilibrium sound speed squared, and we have introduced the $\mathbf{x}$-centred kernel function $K^{\mathbf{x}}(\mathbf{y}) \equiv K(\mathbf{y} - \mathbf{x})$. The right-hand sides of Eqs. \eqref{eq:xiosc} and \eqref{eq:uosc} constitute inhomogeneous forcing terms (see Section \ref{sec:turbulent-effects} for more details), and it was therefore possible to filter out certain negligible contributions. We refer the reader to the details given in Section \ref{app:0th}, where we essentially argue first that the non-stochastic zeroth-order terms in the linearised equations vanish under the unperturbed hydrostatic equilibrium condition, and then that the linear forcing (i.e. the stochastic zeroth-order terms that are linear in the stochastic processes $\bm{\xi}_t$, $\ut$ or $\bm{\eta}$) is negligible. The last point is related to the fact that the turbulent spectrum has most of its power in wavevectors and angular frequencies far removed from those characteristic of the modes, and therefore unable to provide with efficient mode driving.

Formally, Eqs. \eqref{eq:xiosc} and \eqref{eq:uosc} take the form of a linear stochastic, inhomogeneous wave equation in a completely closed form, in the sense that the various terms  on their right-hand side are written as explicit functions of the wave variables $\xiosc$ and $\uosc$ themselves or the turbulent fields $\bm{\xi}_t$ and $\ut$, whose statistical properties are considered known (see hypothesis H4). In writing Eq. \eqref{eq:uosc}, we  split the velocity equation into three components. $\mathbf{L}_1^d$ contains all the terms that are linear in $\xiosc$ and $\uosc$, but do not explicitly contain either stochastic processes $\bm{\xi}_t$, $\ut$, or $\bm{\eta}$. It represents the deterministic contribution to the homogeneous part of the wave equation, and corresponds to the classical propagation of acoustic waves, without any impact from the turbulence. On the other hand, $\mathbf{L}_1^s$ contains all the terms that are linear in $\xiosc$ and $\uosc$ and explicitly depend on $\bm{\xi}_t$, $\ut$, or $\bm{\eta}$. Finally, $\mathbf{L}_0$ contains all the terms that are independent of $\xiosc$ and $\uosc$. The reason for this specific decomposition will become apparent in a moment, when we discuss the physical role played by each term.

\subsection{Effects of turbulence on the wave equation\label{sec:turbulent-effects}}

The last term on the left-hand side of Eq. \eqref{eq:uosc}, together with its right-hand side, contain the contribution of turbulence to the wave equation, which arises from the action of the turbulent fields on the oscillations. We can see that one effect of turbulence is to add the inhomogeneous part $\mathbf{L}_0$ to the wave equation. This part acts as a forcing term, and Eq. \eqref{eq:L0} shows that it corresponds to the fluctuations of the turbulent pressure around its ensemble average. This is in perfect accordance with the widely accepted picture that the stochastic excitation of the global modes of oscillation in solar-like stars is due mainly to quadrupolar turbulent acoustic emission \citep{samadiG01}. Furthermore, we note that we  only kept the contributions to mode excitation that are not linear in the turbulent velocity field as linear contributions turn out to be negligible (see Section \ref{app:0th} for a more developed discussion). We note that the non-linear Lagrangian, turbulent fluctuations of entropy, which is widely recognised as another source of stochastic driving for solar-like $p$-modes, does not arise from the above formalism. The only reason is because we considered a polytropic equation of state from the start, and as such neglected to model entropy fluctuations in both the oscillations and the turbulence.

The second effect of the turbulent fields on the oscillations is to modify the linear part,  that is to say the propagation of the waves. This stochastic correction corresponds to the term $\mathbf{L}_1^s$ defined by Eq. \eqref{eq:L1s}. This term models two effects that are usually studied as distinct phenomena, but are actually intertwined and cannot be considered separately: a shift in the eigenfrequency of the resonant modes of the system (commonly referred to as the modal  or `intrinsic' part of the surface effects), and the absorption,  or damping,  of the energy of the waves as they travel through the turbulent medium. Equation \eqref{eq:L1s} shows that these phenomena arise either from the non-linear advection term in the momentum equation, as represented by the first two terms on its right-hand side, or from the joint effect of turbulent dissipation, buoyancy, and pressure-rate-of-strain correlations, as jointly represented by all the other terms. It is apparent, in particular, that while the former is linear in $\ut$, the latter has a more complicated multipolar decomposition in terms of $\ut$, with first-, second-, and third-order contributions alike. As a whole, the term $\mathbf{L}_1^s$ in the velocity equation plays the same role, for instance, as $\mathcal{D}\left(\mathbf{v}_\text{osc}\right)$ in \citet{samadiG01} (see their Eq. 26).

\subsection{Limiting case: The standard wave equation}

We now explore the limiting case where there is no turbulence, in which case the only term that remains in Eq. \eqref{eq:uosc} is $\mathbf{L}_1^d$. In the absence of turbulence the integrals appearing in Eq. \eqref{eq:L1d} are drastically simplified because, in this limit, the wave variables $\xiosc$ and $\uosc$ are equal to their own ensemble average, that is to say to their own kernel estimate. This allows us to write, for instance
\begin{equation}
    \displaystyle\int \d^3\mathbf{y} ~ \rho_0(\mathbf{y}) \xiosci{i}(\mathbf{y},t) \left.\dfrac{\partial K}{\partial x_j}\right|_{\mathbf{y}-\mathbf{x}} = -\dfrac{\partial \rho_0\xiosci{i}}{\partial x_j}~.
\end{equation}
Ultimately, this leads to the following simplification of Eq. \eqref{eq:L1d}:
\begin{equation}
    L_{1,i}^d = \left[\dfrac{\partial c_0^2}{\partial x_i} - \dfrac{1}{\rho_0}\dfrac{\partial p_0}{\partial x_i}\right]\dfrac{1}{\rho_0}\dfrac{\partial \rho_0 \xiosci{j}}{\partial x_j} + \dfrac{c_0^2}{\rho_0}\dfrac{\partial^2 \rho_0 \xiosci{j}}{\partial x_i \partial x_j}~.
\end{equation}

\noindent Hence from Eqs. \eqref{eq:xiosc} and \eqref{eq:uosc}, which in this limit read
\begin{align}
    & \dfrac{\partial\xiosc}{\partial t} - \uosc = \mathbf{0}~, \\
    & \dfrac{\partial\uosc}{\partial t} - \mathbf{L}_1^d = \mathbf{0}~,\end{align}
we obtain the following homogeneous, second-order wave equation
\begin{equation}
    \left(\dfrac{\partial^2}{\partial t^2} - \mathcal{L}\right)\uosc = \mathbf{0}~,
\end{equation}
with
\begin{equation}
    \mathcal{L}\left(\uosc\right) = \dfrac{1}{\rho_0}\left[ \left(\bm{\nabla}c_0^2 (\bm{\nabla} \cdot \rho_0 \uosc )\right) - \dfrac{1}{\rho_0}\bm{\nabla}p_0 (\bm{\nabla}\cdot\rho_0 \uosc) \right]~.
\end{equation}
We recover the equation for free acoustic oscillations in a stratified medium in its exact form, provided the Cowling approximation is adopted (see hypothesis H5). It corresponds exactly to the homogeneous part of the wave equation derived, for instance by \citet{goldreichK77b} (see their Eq. 16); or, equivalently, to the equation presented in \citet{unno89book} (see their Eqs. 14.2 and 14.3), although these are written in terms of displacement and pressure fluctuation rather than displacement and velocity \citep[see also][their Eq. 16]{samadiG01}. The only exception is the absence of the term depending on the entropy gradient (which does not appear here because the gas is polytropic).

\section{Discussion\label{sec:discussion}}

The linear stochastic wave equation comprised of Eqs. \eqref{eq:xiosc} and \eqref{eq:uosc} was obtained in the scope of a certain number of hypotheses and approximations, whose validity we now discuss. We can split these hypotheses into two categories: those pertaining to the establishment of the stochastic differential equations, and those pertaining to the linearisation of these equations.

All the hypotheses we adopted in the linearisation are itemised  in Section \ref{sec:lin}. Hypotheses (H1) through (H5) are actually of two different natures. Hypotheses (H1) on the smallness of $\uosc$, (H2) on the smallness of $\xiosc$, and (H4) on the absence of back-reaction of the oscillations on the turbulence define the framework in which we performed the linearisation, and are therefore necessary assumptions. On the other hand, hypotheses (H3) on the neglect of $\bm{\nabla} \cdot \left( \rho_0 \bm{\xi}_t \right)$ and (H5) on the neglect of the perturbed gravitational potential are simplifying assumptions that are not necessary, strictly speaking, but help  simplify the formalism considerably. Hypothesis (H5) is a common assumption in the analysis of stellar oscillations: without it, because gravity is an unscreened force acting on long distances, the resulting equations would be highly non-local. As we mention in Section \ref{sec:lin}, its domain of validity is the high-radial-order modes of oscillation, but it is usually adopted throughout the entire oscillation spectrum. Hypothesis (H3), on the other hand, may require some more discussion. As we briefly mention above, it corresponds to the anelastic approximation, and amounts to neglecting the turbulent fluctuations of the fluid density $\rho_t$. Taking these fluctuations into account would require having knowledge of the statistical properties of $\rho_t$, the same way we consider the properties of $\ut$ known. However,  current models of compressible turbulence are not  yet able to fully account for $\rho_t$ without any underlying simplifying assumptions, such as the Boussinesq approximation or the anelastic approximation. It is difficult to assess how sensible to this assumption  the results obtained for the behaviour of turbulent convection are,   and {a fortiori} its coupling with oscillations, but for lack of a more realistic treatment of turbulence compressibility, we nevertheless chose to adopt hypothesis (H3).

We  have also adopted a number of approximations in order to establish the closed system of equations (Eqs. \eqref{eq:GLM_eulx_v0}, \eqref{eq:GLM_eulu_v0}, \eqref{eq:SPH_cont_rho}, \eqref{eq:SPH_cont_u}, \eqref{eq:SPH_cont_rey}, \eqref{eq:SPH_p}, and \eqref{eq:SPH_epsilon}) in Section \ref{sec:stoch-model}. All of them consist in simplifying assumptions, that we adopt not because they are necessary to build the formalism, but because the aim of this paper is to make the basics of this method as clear as possible, rather than adopting the most realistic turbulence model possible. As such, we do not attempt to give a physical justification for the following hypotheses, but instead discuss how they affect the final stochastic wave equation, and how one would go about circumventing these simplifications.

\textbf{(H6)} We consider the flow to be adiabatic, in the sense that the only fluid particle properties that need to be described in the Lagrangian stochastic model are the position and velocity of the particles. In the scope of this hypothesis, the energy equation is replaced with a relation between the mean density and pressure that we chose to be polytropic, without specifying the associated polytropic exponent $\gamma$, which  means that the non-adiabatic effects pertaining to the oscillations are not contained in the formalism presented in this paper. This includes the perturbation of the convective flux and the radiative flux by the oscillations, which are in reality susceptible to affect the damping rate of the modes as well as the surface effects. Avoiding hypothesis (H6) would allow for the inclusion of all non-adiabatic effects in the model. Essentially, adopting a non-adiabatic framework would require an additional  SDE for the internal energy of the fluid particles (or any other alternative thermodynamic variable), thus leading to the introduction of an additional thermodynamic wave variable $e_\text{osc}$, to be linearised around the turbulence-induced energy fluctuations $e_t$. This would then increase the order of the system of equations, and would require the statistical properties of the additional turbulent field $e_t$, including its correlation with $\ut$, to be known.

\textbf{(H7)} We consider that the turbulent frequency $\omega_t$,  defined by Eq. \eqref{eq:SPH_epsilon} as the ratio of the dissipation rate $\epsilon$ to the turbulent kinetic energy $k$,  takes a constant value. The turbulent frequency represents the rate at which $k$ would decay towards zero if there were no production of turbulence whatsoever, and can be interpreted as the inverse lifetime of the energy-containing turbulent eddies. In essence, this amounts to assuming the existence of a single timescale  associated with the entire turbulent cascade, which is at odds with even the simplest picture of turbulence. Avoiding hypothesis (H7) would allow  a much more realistic modelling of the turbulent dissipation and its perturbation by the oscillations, which is likely to play an important role in both mode damping and surface effects. This would require including the turbulent frequency as a fluid particle property, with its own SDE. As for velocity or internal energy (see hypothesis (H6) above), this would lead to the introduction of $\omega_{t,\text{osc}}$ as an additional wave variable, to be linearised around a new turbulent field $\omega_{t,t}$, whose statistical properties would have to be input in the model.

\textbf{(H8)} We consider that the time average of the flow velocity over very long timescale,  in other words the velocity associated with the equilibrium background, is zero. This amounts to neglecting rotation, whether it be global or differential. Taking rotation into account would require either a non-zero $\langle \mathbf{u} \rangle_L$ field to be included, or else a Coriolis inertial force to be added in the velocity SDE.

In summary, hypotheses (H1), (H2), and (H4) are fundamental in building the formalism, and cannot be avoided, but they are also firmly and physically grounded. Hypotheses (H3) and (H5) are simplifying assumptions that are not strictly  necessary, nor as clearly valid, but which are unavoidable given the current state of our capabilities. Finally, hypotheses (H6), (H7), and (H8) are also simplifying assumptions, and are very much invalid; however, we adopted them here to provide  a simple framework serving as a proof of concept for the formalism presented in this paper. In particular, hypotheses (H6) and (H7) must be discarded as soon as possible if  a realistic model of turbulence is to be adopted. This is left to a future work in this series.

\section{Conclusion\label{sec:conclusion}}

In this series of papers we investigate Lagrangian stochastic models of turbulence as a rigorous way of modelling the various phenomena arising from the interaction between the highly turbulent motions of the gas at the top of the convection zone in solar-like stars and the global acoustic modes of oscillation developing in these stars. These include the stochastic excitation of the modes, their stochastic damping, and the turbulence-induced shift in their frequency  called surface effects.

In this first paper we presented a very simple polytropic Lagrangian stochastic turbulence model, serving as a proof of concept for the novel method presented here, and we showed how it can be used to derive stochastic differential equations (SDEs) governing the evolution of Eulerian fluid variables relevant to the study of oscillations. We then linearised these SDEs to obtain a linear stochastic wave equation containing, in the most self-consistent way possible, the terms arising from the turbulence-oscillation coupling. This wave equation correctly reduces to the classical propagation of free acoustic waves in a stratified medium in the limit where turbulence is neglected. It also exactly models the stochastic forcing term due to turbulent acoustic emission, arising from coherent fluctuations in the turbulent pressure. In addition, the resulting stochastic wave equation contains the turbulent-induced correction to the linear operator governing the propagation of the waves, thus allowing for the modelling of both mode damping and modal surface effects. The method presented here offers multiple, key advantages:
\begin{itemize}
    \item[$\bullet$] At no point does it require  separating the equations of the flow into a turbulent equation and an oscillation equation, thus allowing the turbulent contribution to naturally and consistently arise in the wave equation. Instead, we leave the statistical properties of the turbulence as known oscillation-independent inputs to the model.
    \item[$\bullet$] All aspects of turbulence-oscillation interaction are modelled simultaneously, within the same stochastic wave equation, thus shedding a more consistent light on these intertwined phenomena.
    \item[$\bullet$] This method completely circumvents the need to adopt the mixing-length hypothesis, which is crucial as this hypothesis is both almost inescapable in current convection modelling, and very invalid close to the radiative-convective transition zone. The reason we do not need to adopt this assumption stems from the fact that the starting model is at particle level, where equations are much easier to close.
    \item[$\bullet$] The parameters appearing in Lagrangian stochastic models are much more easily linked to the underlying physical assumptions, and therefore easier to constrain, with the help of 3D hydrodynamic simulations. They are also more firmly physically grounded.
    \item[$\bullet$] In addition, this formalism applies to radial and non-radial oscillations alike.
\end{itemize}

However, this paper only constitutes a first step. In the following paper in this series we will show how such a stochastic wave equation can be used to yield a set of stochastic differential equations governing the temporal evolution of the complex amplitude of the modes. These simplified amplitude equations \citep{stratonovich65} are much more practical for the study of turbulence-oscillation coupling, and in particular explicitly and simultaneously yield the excitation rates of the modes, their lifetimes, as well as their turbulence-induced frequency corrections. Finally, as we mentioned in Section \ref{sec:discussion}, extending this work to a non-adiabatic model (i.e. discarding hypothesis H6), and with a more realistic treatment of eddy lifetimes (i.e. discarding hypothesis H7), constitutes an essential and unavoidable step to apply this formalism to the actual stellar case, and will be the subject of a subsequent paper.

\begin{acknowledgements}
The authors wish to thank the anonymous referee for his/her insightful comments, which helped improve the clarity and quality of this manuscript.
\end{acknowledgements}

\bibliographystyle{aa}
\bibliography{biblio}

\begin{appendix}

\section{The equivalent Reynolds stress model\label{app:reynolds-stress}}
The procedure leading from a given Lagrangian stochastic model to the equivalent Reynolds stress model (i.e. the corresponding transport equations for the first- and second-order moments of the flow velocity) can be found, for instance, in \citet[][Chap. 12]{pope00book}. For the generalised Langevin model considered in this paper, it yields
\begin{equation}
    \dfrac{D \overline{\rho}}{D t} + \overline{\rho}\dfrac{\partial \widetilde{u_i}}{\partial x_i} = 0~,
    \label{eq:mean-rho}
\end{equation}
\begin{equation}
    \dfrac{D \widetilde{u_i}}{D t} + \dfrac{1}{\overline{\rho}}\dfrac{\partial \overline{\rho} \widetilde{u''_i u''_j}}{\partial x_j} = -\dfrac{1}{\overline{\rho}} \dfrac{\partial \overline{p}}{\partial x_i} + g_i~,
    \label{eq:mean-u}
\end{equation}
and
\begin{multline}
    \dfrac{D \widetilde{u''_i u''_j}}{D t} + \dfrac{1}{\overline{\rho}}\dfrac{\partial \overline{\rho} \reallywidetilde{u''_i u''_j u''_k}}{\partial x_k} = -\widetilde{u''_iu''_j}\dfrac{\partial \widetilde{u_k}}{\partial x_k} - \widetilde{u''_iu''_k}\dfrac{\partial \widetilde{u_j}}{\partial x_k} - \widetilde{u''_ku''_j}\dfrac{\partial \widetilde{u_i}}{\partial x_k} \\
    + G_{ik} \widetilde{u_j'' u_k''} + G_{jk} \widetilde{u_i'' u_k''} + C_0 \epsilon \delta_{ij}~,
    \label{eq:mean-uu}
\end{multline}
where $\delta_{ij}$ denotes the Kronecker symbol, and we have introduced the pseudo-Lagrangian particle derivative $D / Dt \equiv \partial_t + \widetilde{u_i}\partial_i$.

Equation \eqref{eq:mean-rho} yields the continuity equation in its exact form, without having to include an evolution equation for density at particle level. This is due to the fact that particle positions are advanced through time using their own individual velocities;  since each particle carries its own unchanging mass, then by construction there can be no local mass loss or gain.

Equation \eqref{eq:mean-u} also yields the mean momentum equation in its exact form, primarily because the mean force in the stochastic model is already included in its exact form from the start. We note, however, that the transport term (i.e. the second term on the left-hand side of Eq. \ref{eq:mean-u}) is also modelled exactly, even though it is not explicitly included in any way in Eqs. \eqref{eq:GLMx} and \eqref{eq:GLMu}. This is, once again, because of the Lagrangian nature of the stochastic model, and is incidentally one of its most interesting features: all advection terms are implicitely and exactly modelled because trajectories integrated through Eqs. \eqref{eq:GLMx} and \eqref{eq:GLMu} coincide with actual fluid particle trajectories.

 Equation \eqref{eq:mean-uu}  differs slightly from the exact Reynolds stress equation derived directly from the Navier-Stokes equation, which reads
\begin{multline}
    \dfrac{D \widetilde{u''_i u''_j}}{D t} + \dfrac{1}{\overline{\rho}}\dfrac{\partial \overline{\rho} \reallywidetilde{u''_i u''_j u''_k}}{\partial x_k} = -\widetilde{u''_iu''_j}\dfrac{\partial \widetilde{u_k}}{\partial x_k} - \widetilde{u''_iu''_k}\dfrac{\partial \widetilde{u_j}}{\partial x_k} - \widetilde{u''_ku''_j}\dfrac{\partial \widetilde{u_i}}{\partial x_k} \\
    \mathrm{sym}\left(-\dfrac{\overline{u''_i}}{\overline{\rho}}\dfrac{\partial \overline{p}}{\partial x_j} - \dfrac{1}{\overline{\rho}}\overline{u''_i\dfrac{\partial p'}{\partial x_j}} - \epsilon_{ij}\right)~,
    \label{eq:exact-uu}
\end{multline}
where `sym' refers to the symmetric part of the tensor inside the brackets. Several contributions are still modelled in their exact form, including the transport term (the second term on the left-hand side), but also the production term (the first three terms on the right-hand side). However, the last three terms are not modelled exactly, and are (from left to right) the buoyancy contribution, the pressure-rate-of-strain tensor, and the dissipation tensor. Comparing Eqs. \eqref{eq:mean-uu} and \eqref{eq:exact-uu}, it can be seen that these contributions are collectively modelled by the last two terms in Eq. \eqref{eq:GLMu}, which correspond to the fluctuating part of the force acting upon the fluid particle. More specifically, we obtain
\begin{equation}
    \mathrm{sym}\left(-\dfrac{\overline{u''_i}}{\overline{\rho}}\dfrac{\partial \overline{p}}{\partial x_i} - \dfrac{1}{\overline{\rho}}\overline{u''_i\dfrac{\partial p'}{\partial x_i}} - \epsilon_{ij}\right) = G_{ik} \widetilde{u_j'' u_k''} + G_{jk} \widetilde{u_i'' u_k''} + C_0 \epsilon \delta_{ij}~.
    \label{eq:appA_temp}
\end{equation}
This equation is more readily interpreted if we remember that, in the high Reynolds number limit, the dissipation tensor is isotropic, which allows for the definition of the scalar dissipation $\epsilon$ appearing in the stochastic model:
\begin{equation}
    \epsilon_{ij} \equiv \dfrac{2}{3}\epsilon \delta_{ij}~.
\end{equation}
Furthermore, the drift tensor is usually decomposed into an isotropic and anisotropic part, according to
\begin{equation}
    G_{ij} = -\left(\dfrac{1}{2} + \dfrac{3}{4}C_0\right)\dfrac{\epsilon}{k}\delta_{ij} + G_{ij}^a~,
\end{equation}
where $k \equiv \widetilde{u_i''u_i''} / 2$ is the turbulent kinetic energy. This decomposition ensures that, in the special case of incompressible, homogeneous, isotropic turbulence, if we take $G_{ij}^a = 0$, the evolution of the Reynolds stress tensor reduces to the exact, analytical solution.

Equation \eqref{eq:appA_temp} can be rearranged to yield
\begin{multline}
    \mathrm{sym}\left(-\dfrac{\overline{u''_i}}{\overline{\rho}}\dfrac{\partial \overline{p}}{\partial x_i} - \dfrac{1}{\overline{\rho}}\overline{u''_i\dfrac{\partial p'}{\partial x_i}}\right) = G_{ik}^a \widetilde{u_j'' u_k''} + G_{jk}^a \widetilde{u_i'' u_k''} \\
    - \left(1 + \dfrac{3}{2}C_0\right) \dfrac{\epsilon}{k} \left(\widetilde{u_i'' u_j''} - \dfrac{2}{3} k \delta_{ij}\right)~.
    \label{eq:corresp-Rey-stoch-uu}
\end{multline}
Equation \eqref{eq:corresp-Rey-stoch-uu} allows us to interpret the collective effect of buoyancy and pressure-rate-of-strain correlation on the evolution of the Reynolds stresses. First, all the terms on the right-hand side are traceless, which means that they only have a redistributive role;  they redistribute energy among the different components of the Reynolds stress tensor, without ever resulting in a net loss or gain of energy. By contrast, it is the scalar dissipation $\epsilon$ which is responsible for the decay of kinetic turbulent energy,  an effect that is only counterbalanced by the shear- and compression-induced production term (i.e. the first three terms on the right-hand side of Eq. \eqref{eq:exact-uu}).

Furthermore, it is readily seen that the last term on the right-hand side of Eq. \eqref{eq:corresp-Rey-stoch-uu} tends to isotropise the Reynolds stress tensor since for isotropic turbulence we would precisely have $\widetilde{u_i'' u_j''} = 2k\delta_{ij} / 3$. The rate at which this term makes the Reynolds stress decay towards isotropy is equal to $(1 + 3C_0 / 2)\omega_t$, where $\omega_t$ is the turbulent dissipation rate defined by Eq. \eqref{eq:SPH_epsilon}. On the other hand, the other two terms on the right-hand side of Eq. \eqref{eq:corresp-Rey-stoch-uu} create anisotropy in the Reynolds stress tensor, and we can intuitively understand that the anisotropy of the stationary Reynolds stress  results from a balance between these two effects.

\section{A detailed derivation for the Lagrangian-to-Eulerian change of variables\label{app:identity-lagrangian}}

The goal of this appendix is to provide  a detailed derivation of the various steps in the procedure described in Section \ref{sec:lag-to-eul}. We first derive the general identity \eqref{eq:identity}, which is valid for any fluid quantity; we then apply this general identity to the displacement and velocity variables.

\subsection{Derivation of identity \eqref{eq:identity}\label{app:identity-part1}}

Let us consider, for the moment, that the function $\bm{\xi}(\mathbf{x},t)$ is an arbitrary function of space and time, which we do not specify at first. We recall the following notations
\begin{align}
    & \mathbf{X}(\mathbf{x},t) \equiv \mathbf{x} + \bm{\xi}(\mathbf{x},t)~, \\
    & \phi_L(\mathbf{x},t) \equiv \phi(\mathbf{X}(\mathbf{x},t), t)~,
\end{align}
where $\phi$ is an arbitrary quantity. The usual chain rules for derivation then yield
\begin{align}
    & \dfrac{\partial (\phi_L)}{\partial t} = \left(\dfrac{\partial\phi}{\partial t}\right)_L + \dfrac{\partial X_i}{\partial t}\left(\dfrac{\partial \phi}{\partial x_i}\right)_L~, \label{eq:appB-PhiL-time-derivative} \\
    & \dfrac{\partial (\phi_L)}{\partial x_i} = \dfrac{\partial X_j}{\partial x_i}\left(\dfrac{\partial \phi}{\partial x_j}\right)_L~.
    \label{eq:appB-PhiL-space-derivative}
\end{align}
If the function $\mathbf{x} \mapsto \mathbf{X}(\mathbf{x},t)$ (the time $t$ being fixed) is bijective, then for any velocity field $\mathbf{u}(\mathbf{x},t)$, there necessarily exists an associated field $\mathbf{V}(\mathbf{x},t)$ such that, were  point $\mathbf{x}$ to move with velocity $\mathbf{V}$,  point $\mathbf{X}$ would then move with the actual fluid velocity $\mathbf{u}_L$. Otherwise stated, $\mathbf{V}$ corresponds to the advective velocity in the material derivative of $\mathbf{X}$, so that
\begin{equation}
    \dfrac{\partial X_i}{\partial t} + V_j\dfrac{\partial X_i}{\partial x_j} = u_{i,L}~.
    \label{eq:appB-Xderivative}
\end{equation}
Plugging Eq. \eqref{eq:appB-Xderivative} into Eq. \eqref{eq:appB-PhiL-time-derivative}, we find
\begin{align}
    \dfrac{\partial (\phi_L)}{\partial t}
    & = \left(\dfrac{\partial \phi}{\partial t}\right)_L + \left(u_{i,L} - V_j\dfrac{\partial X_i}{\partial x_j}\right)\left(\dfrac{\partial \phi}{\partial x_i}\right)_L \nonumber \\
    & = \left(\dfrac{D\phi}{D t}\right)_L - V_j\dfrac{\partial X_i}{\partial x_j}\left(\dfrac{\partial \phi}{\partial x_i}\right)_L~,
\end{align}
where $D / D t \equiv \partial_t + u_i \partial_i$. In turn, plugging Eq. \eqref{eq:appB-PhiL-space-derivative}, this transforms into
\begin{equation}
    \dfrac{\partial (\phi_L)}{\partial t} = \left(\dfrac{D \phi}{D t}\right)_L - V_j\dfrac{\partial (\phi_L)}{\partial x_j}~,
\end{equation}
thus yielding the required identity
\begin{equation}
    \left(\dfrac{D \phi}{D t}\right)_L = \langle D \rangle_L(\phi_L)~,
    \label{eq:appB-temp-identity}
\end{equation}
where $\langle D \rangle_L \equiv \partial_t + V_i \partial_i$.

This is all valid regardless of the definition of the function $\bm{\xi}(\mathbf{x},t)$. However, let us now consider that $\mathbf{x}$ does actually correspond to a mean\footnote{We recall  that throughout this discussion, the word `mean' refers to the as-yet-unspecified averaging process $\langle . \rangle$ (see the main body of the paper for more details).} position, and that the function $\bm{\xi}(\mathbf{x},t)$ actually denotes the fluctuating fluid displacement around this mean position $\mathbf{x}$. Then by construction, we have
\begin{equation}
    \langle \xi_i \rangle = 0~.
\end{equation}
Point $\mathbf{x}$ now corresponding to a mean position, the velocity $\mathbf{V}$ at which it is displaced must itself be a mean quantity, so that
\begin{equation}
    \langle \mathbf{V} \rangle = \mathbf{V}~.
\end{equation}
Let us now apply the mean operator $\langle . \rangle$ to Eq. \eqref{eq:appB-Xderivative}; since $V_j$ can be pulled out of the mean, we obtain
\begin{align}
    \langle u_i \rangle_L
    & = \dfrac{\partial \langle X_i \rangle}{\partial t} + V_j \dfrac{\partial \langle X_i \rangle}{\partial x_j} \nonumber \\
    & = \dfrac{\partial x_i}{\partial t} + \dfrac{\partial \langle \xi_i \rangle}{\partial t} + V_j \left( \dfrac{\partial x_i}{\partial x_j} + \dfrac{\partial \langle \xi_i \rangle}{\partial x_j} \right) \nonumber \\    & = 0 + 0 + V_j\left(\delta_{ij} + 0 \right)~,
\end{align}
or in other words
\begin{equation}
    \mathbf{V} = \langle \mathbf{u} \rangle_L~.
\end{equation}

To summarise, the identity \eqref{eq:appB-temp-identity} is always verified, but in general, the velocity $\mathbf{V}$ appearing in the definition of the operator $\langle D \rangle_L$ is not easily specified. Only when the displacement function $\bm{\xi}$ is judiciously defined as a fluctuating particle displacement does the velocity $\mathbf{V}$ reduce to the mean Lagrangian velocity $\langle \mathbf{u} \rangle_L$.

\subsection{Derivation of Eqs. \eqref{eq:GLM_eulx} and \eqref{eq:lag-to-eulu}\label{app:identity-part2}}

Let us apply Eq. \eqref{eq:appB-temp-identity} to $\phi = x_i$ and $\phi = u_i$ alternatively. First, if $\phi = x_i$, then $\phi_L = X_i$, and Eq. \eqref{eq:appB-temp-identity} becomes
\begin{equation}
    \left(\dfrac{\partial x_i}{\partial t} \right)_L + u_{j,L}\left(\dfrac{\partial x_i}{\partial x_j}\right)_L = \dfrac{\partial X_i}{\partial t} + \langle u_j \rangle_L \dfrac{\partial X_i}{\partial x_j}~.
\end{equation}
The mean position $\mathbf{x}$ having no explicit time dependence, we have $(\partial x_i / \partial t)_L = 0$, $(\partial x_i / \partial x_j)_L = \delta_{ij}$, $\partial X_i / \partial t = \partial \xi_i / \partial t$, and $\partial X_i / \partial x_j = \delta_{ij} + \partial \xi_i / \partial x_j$. The above equation then becomes
\begin{equation}
    u_{i,L} = \dfrac{\partial \xi_i}{\partial t} + \langle u_j \rangle_L\left(\delta_{ij} + \dfrac{\partial \xi_i}{\partial x_j} \right)~,
\end{equation}
thus immediately yielding Eq. \eqref{eq:GLM_eulx}.

Secondly, if $\phi = u_i$, then $\phi_L = u_{i,L}$, and Eq. \eqref{eq:appB-temp-identity} becomes
\begin{equation}
    \left(\dfrac{\partial u_i}{\partial t} \right)_L + u_{j,L}\left(\dfrac{\partial u_i}{\partial x_j}\right)_L = \dfrac{\partial (u_{i,L})}{\partial t} + \langle u_j \rangle_L \dfrac{\partial (u_{i,L})}{\partial x_j}~.
    \label{eq:appB-temp-velocity-application}
\end{equation}
But Eq. \eqref{eq:appB-PhiL-space-derivative} allows us to write
\begin{equation}
    \dfrac{\partial (u_{i,L})}{\partial x_j} = \dfrac{\partial X_k}{\partial x_j} \left(\dfrac{\partial u_i}{\partial x_k}\right)_L~,
\end{equation}
where we recall that
\begin{equation}
    \dfrac{\partial X_k}{\partial x_j} = \delta_{kj} + \dfrac{\partial \xi_k}{\partial x_j}~.
\end{equation}
Plugging these into Eq. \eqref{eq:appB-temp-velocity-application}, we find
\begin{equation}
    \left(\dfrac{\partial u_i}{\partial t} \right)_L + u_{j,L}\left(\dfrac{\partial u_i}{\partial x_j}\right)_L = \dfrac{\partial (u_{i,L})}{\partial t} + \langle u_j \rangle_L \left(\delta_{jk} + \dfrac{\partial \xi_k}{\partial x_j} \right) \left(\dfrac{\partial u_i}{\partial x_k}\right)_L~.
\end{equation}
Isolating the first term on the right-hand side yields Eq. \eqref{eq:lag-to-eulu}.

\section{The insignificance of the back-reaction of the oscillations on the turbulence\label{app:backreaction}}

Let us formally write the governing equations of the flow in the following abstract form
\begin{equation}
    \dfrac{\partial U}{\partial t} + \mathcal{L}(U) + \mathcal{B}(U,U) = 0~,
    \label{eq:formal-eq}
\end{equation}
where $U$ represents the flow variables, $\mathcal{L}$ is a linear operator, and $\mathcal{B}$ a bilinear operator containing the advection terms. In the limit of small amplitudes, which are relevant for solar-like oscillations, the wave variables can be expanded as
\begin{equation}
    U = U_0 + aU_1 + a^2U_2~,
    \label{eq:formal-expansion}
\end{equation}
where $a$ is small ordering parameter. Plugging Eq. \eqref{eq:formal-expansion} into Eq. \eqref{eq:formal-eq} and isolating the various orders in $a$, we obtain the following hierarchy of equations
\begin{align}
    & \dfrac{\partial U_0}{\partial t} + \mathcal{L}(U_0) + \mathcal{B}(U_0,U_0) = 0~, \\
    & \dfrac{\partial U_1}{\partial t} + \mathcal{L}(U_1) + \mathcal{B}(U_0,U_1) + \mathcal{B}(U_1,U_0) = 0~, \label{eq:formal-lin} \\
    & \dfrac{\partial U_2}{\partial t} + \mathcal{L}(U_2) + \mathcal{B}(U_0,U_2) + \mathcal{B}(U_2,U_0) = -\mathcal{B}(U_1,U_1)~, \label{eq:formal-backreac}
\end{align}
where the first equation governs the basic flow, the second equation governs the waves, and the third equation governs the back-reaction of the waves on the basic flow. In particular, Eq. \eqref{eq:formal-backreac} takes the form of a forced linear wave, where the linear part $\mathcal{L}' \equiv \mathcal{L} + \mathcal{B}(U_0,.) + \mathcal{B}(.,U_0)$ is identical to the linear part in the actual wave equation \eqref{eq:formal-lin}, and the forcing term is given by the right-hand side of Eq. \eqref{eq:formal-backreac}. Because $\mathcal{L}'$ is common to both Eqs. \eqref{eq:formal-lin} and \eqref{eq:formal-backreac}, if we denote the angular frequency of the wave as $\omega$, we can write the homogeneous solution of Eq. \eqref{eq:formal-backreac} as 
\begin{equation}
    U_{2,h}(t) = A \exp^{j\omega t}~,
\end{equation}
and the total solution (including the forcing, inhomogeneous part) formally reads
\begin{equation}
    U_2(t) = -\displaystyle\int_0^t \d t' ~ \exp^{-j\omega(t'-t)} \mathcal{B}\left(U_1(t'),U_1(t')\right)~.
\end{equation}
The back-reaction of the waves on the turbulence is therefore driven by its resonance with the non-linear oscillation-induced advection. In the limit $t \rightarrow +\infty$, this formal solution yields
\begin{equation}
    \left|U_2(t)\right|^2 \sim \left|\vphantom{\dfrac{1}{1}} \mathrm{TF}\left[\mathcal{B}(U_1,U_1)\right](\omega)\right|^2~,
\end{equation}
where `TF' denotes the Fourier transform. But $U_1$ refers to the wave, so that its Fourier spectrum only has power around the angular frequency $\omega$ of the wave. In turn, this means that the quadratic operator $\mathcal{B}$ applied to the velocity $U_1$ has a Fourier spectrum whose power is concentrated around $\omega = 0$ (i.e. the continuous component) as well as $2\omega$ (twice the frequency of the waves). By contrast, it contains little to no power around the actual frequency $\omega$ of the oscillation, which justifies that the impact of the back-reaction $U_2(t)$ on the mean flow $U_0(t)$ may be neglected.

\section{Derivation of the linear wave equation \label{app:lin}}

In this appendix we linearise the system comprised of Eqs. \eqref{eq:GLM_eulx_v0}, \eqref{eq:GLM_eulu_v0}, \eqref{eq:SPH_cont_rho}, \eqref{eq:SPH_cont_u}, \eqref{eq:SPH_cont_rey}, \eqref{eq:SPH_p}, and \eqref{eq:SPH_epsilon}, using the hypotheses outlined in Section \ref{sec:lin}. We start, in Section \ref{app:lin-SPH}, by linearising all the ensemble averages described in the SPH formalism (i.e. Eqs. \eqref{eq:SPH_cont_rho}, \eqref{eq:SPH_cont_u}, \eqref{eq:SPH_cont_rey}, \eqref{eq:SPH_p}, and \eqref{eq:SPH_epsilon}). In Section \ref{app:lin-total}, we then plug these linearised ensemble averages to derive the linearised version of Eqs. \eqref{eq:GLM_eulx_v0} and \eqref{eq:GLM_eulu_v0}. Finally, in Section \ref{app:0th}, we discuss which terms should be retained in the inhomogeneous forcing term of the resulting wave equation. For more clarity in the notations, we dropped all dependence on the space variable $\mathbf{x}$, the space variable $\mathbf{y}$ used inside the integrals, and  time $t$. It must be understood that all the quantities outside the integrals depend on $\mathbf{x}$ and $t$, and all quantities inside depend on $\mathbf{y}$ and $t$.

\subsection{Linearising the mean fields\label{app:lin-SPH}}

A general remark can be made beforehand concerning all ensemble averages described in the SPH formalism: {the occurrence of $\bm{\xi_t}$ vanishes completely from their linearised version by virtue of hypothesis (H3)}. We have already shown, in the main body of this paper, that this is the case for the mean density $\overline{\rho}$,  but this is also the case for the mean velocity and Reynolds stress tensor.  They can both formally be written as
\begin{equation}
    \widetilde{Q} = \dfrac{1}{\overline{\rho}} \displaystyle\int \d^3\mathbf{y} ~ \rho_0 Q(\mathbf{y} + \bm{\xi}) K^{\mathbf{x}}(\mathbf{y} + \bm{\xi})~,
\end{equation}
where $Q$ is a function of velocity only ($Q = \mathbf{u}$ for the mean velocity, and $Q = \left(u_i - \widetilde{u_i}\right)\left(u_j - \widetilde{u_j}\right)$ for the Reynolds stress tensor), and we have introduced the $\mathbf{x}$-centred kernel function $K^{\mathbf{x}}(\mathbf{y}) \equiv K(\mathbf{y}-\mathbf{x})$. Because $Q$ only depends on the velocity variable $\mathbf{u}$, and not on the displacement variable $\bm{\xi}$, the only occurrence of $\bm{\xi_t}$ in the linearisation of $\widetilde{Q}$ comes from the term
\begin{equation}
    \widetilde{Q} = [...] + \dfrac{1}{\overline{\rho}} \displaystyle\int \d^3\mathbf{y} ~ \rho_0 \bm{\xi}_t \cdot \bm{\nabla}\left(Q K^{\mathbf{x}} \right)~.
\end{equation}
Performing an integration by part yields
\begin{equation}
    \widetilde{Q} = [...] - \dfrac{1}{\overline{\rho}} \displaystyle\int \d^3\mathbf{y} ~ Q K^{\mathbf{x}} \bm{\nabla} \cdot \left(\rho_0 \bm{\xi}_t \right)~,
\end{equation}
where the surface term vanishes because of the compact support of the kernel function $K^{\mathbf{x}}$. By virtue of hypothesis (H3), the quantity $\bm{\nabla} \cdot (\rho_0 \bm{\xi_t})$ is negligible, and therefore this contribution can be safely discarded.

As we have just shown, this is true of the mean density, mean velocity, and Reynolds stress tensor. In turn, this is also true of the gas pressure $\overline{p}$ (because it is given as a function of the mean density), as well as the turbulent kinetic energy $k$ and the turbulent dissipation rate $\epsilon$ (because they are both given as a function of the Reynolds stress tensor). Therefore, $\bm{\xi}_t$ can indeed be neglected in the linearised version of every single ensemble average appearing in Eqs. \eqref{eq:GLM_eulx_v0} and \eqref{eq:GLM_eulu_v0}.

\subsubsection{Mean density}

Using hypotheses (H2) and (H3), Eq. \eqref{eq:SPH_cont_rho} can be linearised as
\begin{equation}
    \overline{\rho} = \displaystyle\int \d^3\mathbf{y} ~ \rho_0 \left[K^{\mathbf{x}} + \xiosci{i} \partial_i K^{\mathbf{x}} \right]~.
\end{equation}
The first term on the right-hand side corresponds to the kernel estimator of the equilibrium density $\rho_0$ at $\mathbf{x}$, and therefore represents the ensemble average of $\rho_0$ at $\mathbf{x}$. Since $\rho_0$ is already an equilibrium quantity, it is equal to its own ensemble average, and this term reduces to $\rho_0(\mathbf{x})$ itself. Finally,
\begin{equation}
    \overline{\rho} = \rho_0 + \rho_1~,
\end{equation}
with
\begin{equation}
    \rho_1 = \displaystyle\int \d^3\mathbf{y} ~ \rho_0 \xiosci{i} \partial_i K^{\mathbf{x}}~.
    \label{eq:appD-rho1}
\end{equation}

\subsubsection{Mean gas pressure}

The fluctuating mean density $\rho_1$ is much lower than the equilibrium density $\rho_0$ on account of hypothesis (H2). Therefore, Eq. \eqref{eq:SPH_p} can be linearised, immediately yielding
\begin{equation}
    \overline{p} = p_0 + p_1 ~,
\end{equation}
with
\begin{equation}
    p_1 = \dfrac{p_0\gamma}{\rho_0} \rho_1 \equiv c_0^2 \rho_1~,
    \label{eq:appD-p1}
\end{equation}
where $c_0^2$ is the equilibrium sound speed squared.

\subsubsection{Mean velocity}

Using hypotheses (H1), (H2), and (H3), Eq. \eqref{eq:SPH_cont_u} can be linearised as
\begin{equation}
    \begin{array}{ll}
        \widetilde{\mathbf{u}} =& \dfrac{1}{\rho_0} \displaystyle\int \d^3\mathbf{y} ~ \rho_0 \ut K^{\mathbf{x}} - \dfrac{\rho_1}{\rho_0^2} \displaystyle\int \d^3\mathbf{y} ~ \rho_0 \ut K^{\mathbf{x}} \\
        \vspace{0.05cm} \\
        & + \dfrac{1}{\rho_0}\displaystyle\int \d^3\mathbf{y} ~ \rho_0 \uosc K^{\mathbf{x}} + \dfrac{1}{\rho_0}\displaystyle\int \d^3\mathbf{y} ~ \rho_0 K^{\mathbf{x}} \xiosci{i} \partial_i \ut \\
        \vspace{0.05cm} \\
        & + \dfrac{1}{\rho_0}\displaystyle\int \d^3\mathbf{y} ~ \rho_0 \ut \xiosci{i} \partial_i K^{\mathbf{x}}~,
    \end{array}
    \label{eq:appD-tempu}
\end{equation}
where $\rho_1$ is given by Eq. \eqref{eq:appD-rho1}. This expression can be simplified by remarking that
\begin{equation}
    \displaystyle\int \d^3\mathbf{y} ~ \rho_0 \ut K^{\mathbf{x}} = \overline{\rho} \widetilde{\ut}~,
    \label{eq:appD-utmean}
\end{equation}
because kernel averages represent ensemble averages. Since mass is locally conserved by the turbulent velocity field (i.e. the upflows carry as much mass upwards as the turbulent downdrafts carry downwards), it immediately follows that $\widetilde{\ut} = \mathbf{0}$, and therefore the first two terms in Eq. \eqref{eq:appD-tempu} vanish. Rearranging the remaining terms, we obtain
\begin{equation}
    \widetilde{\mathbf{u}} = \widetilde{\mathbf{u}}_0 + \widetilde{\mathbf{u}}_1~,
\end{equation}
where $\widetilde{\mathbf{u}}_0 = \mathbf{0}$ and
\begin{equation}
    \widetilde{\mathbf{u}}_1 = \dfrac{1}{\rho_0}\displaystyle\int \d^3\mathbf{y} ~ \rho_0 \uosc K^{\mathbf{x}} + \dfrac{1}{\rho_0} \displaystyle\int \d^3\mathbf{y} ~ \rho_0 \xiosci{i} \partial_i \left(\ut K^{\mathbf{x}}\right)~.
    \label{eq:appD-u1}
\end{equation}

\subsubsection{Mean shear tensor}

We  also need to express the linearised shear tensor $\partial_i \widetilde{u_j}$ because this quantity appears in the drift tensor $G_{ij}$. Differentiating Eq. \eqref{eq:appD-u1} with respect to $x_i$, and noting that $\bm{\nabla}_{\mathbf{x}}(K^{\mathbf{x}}(\mathbf{y})) = -\bm{\nabla}_{\mathbf{y}}(K^{\mathbf{x}}(\mathbf{y}))$ (because we considered an isotropic kernel function), we obtain
\begin{equation}
    \partial_i \widetilde{u_j} = (\partial_i \widetilde{u_j})_0 + (\partial_i \widetilde{u_j})_1~,
\end{equation}
where $(\partial_i \widetilde{u_j})_0 = 0$ and
\begin{align}
    (\partial_i \widetilde{u_j})_1 &= - \dfrac{1}{\rho_0} \displaystyle\int \d^3\mathbf{y} ~ \rho_0 \uosci{j} \partial_i K^{\mathbf{x}} - \dfrac{1}{\rho_0} \displaystyle\int \d^3\mathbf{y} ~ \rho_0 \xiosci{k} \partial_k \uti{j} \partial_i K^{\mathbf{x}} \nonumber \\
    & - \dfrac{1}{\rho_0^2} \dfrac{\partial \rho_0}{\partial x_i} \displaystyle\int \d^3\mathbf{y} ~ \rho_0 \uosci{j} K^{\mathbf{x}} \nonumber \\
    & - \dfrac{1}{\rho_0^2} \dfrac{\partial \rho_0}{\partial x_i} \displaystyle\int \d^3\mathbf{y} ~ \rho_0 \xiosci{k} \partial_k \left(\uti{j} K^{\mathbf{x}}\right)~.
    \label{eq:appD-shear1}
\end{align}

\subsubsection{Reynolds stress tensor}

The linearised Reynolds stress tensor is obtain from Eq. \eqref{eq:SPH_cont_rey}, using hypotheses (H1), (H2), and (H3). We find
\begin{equation}
    \begin{array}{ll}
        \widetilde{u_i'' u_j''} =& \dfrac{1}{\rho_0} \displaystyle\int \d^3\mathbf{y} ~ \rho_0 \uti{i} \uti{j} K^{\mathbf{x}} - \dfrac{\rho_1}{\rho_0^2} \displaystyle\int \d^3\mathbf{y} ~ \rho_0 \uti{i} \uti{j} K^{\mathbf{x}} \\
        \vspace{0.05cm} \\
        & + \dfrac{1}{\rho_0} \displaystyle\int \d^3\mathbf{y} ~ \rho_0 \uti{i} \xiosci{k} \left(\partial_k \uti{j}\right)  K^{\mathbf{x}} \\
        \vspace{0.05cm} \\
        & + \dfrac{1}{\rho_0} \displaystyle\int \d^3\mathbf{y} ~ \rho_0 \uti{j} \xiosci{k} \left(\partial_k \uti{i}\right)  K^{\mathbf{x}} \\
        \vspace{0.05cm} \\
        & + \dfrac{1}{\rho_0} \displaystyle\int \d^3\mathbf{y} ~ \rho_0 \uti{i} \uti{j} \xiosci{k} \partial_k K^{\mathbf{x}} \\
        \vspace{0.05cm} \\
        & - \dfrac{1}{\rho_0} \displaystyle\int \d^3\mathbf{y} ~ \rho_0 \uti{i} \uosci{j} K^{\mathbf{x}} + \dfrac{1}{\rho_0} \displaystyle\int \d^3\mathbf{y} ~ \rho_0 \uti{j} \uosci{i} K^{\mathbf{x}} \\
        \vspace{0.05cm} \\
        & - \dfrac{1}{\rho_0} \displaystyle\int \d^3\mathbf{y} ~ \rho_0 \uti{i} \widetilde{u_j}_1 K^{\mathbf{x}} + \dfrac{1}{\rho_0} \displaystyle\int \d^3\mathbf{y} ~ \rho_0 \uti{j} \widetilde{u_i}_1 K^{\mathbf{x}}~,
    \end{array}
\end{equation}
where $\rho_1$ is given by Eq. \eqref{eq:appD-rho1} and $\widetilde{\mathbf{u}}_1$ by Eq. \eqref{eq:appD-u1}. In the last two terms, $\widetilde{u_i}_1$ can be pulled from the integral. The kernel estimator is a representation of ensemble averages, and $\widetilde{u_i}_1$ is already an ensemble average. Once this quantity is pulled out, we recognise the same integral defined by Eq. \eqref{eq:appD-utmean}, meaning that these terms vanish. Additionally, the third, fourth, and fifth terms can be conveniently merged together, and $\rho_1$ can be replaced by its explicit expression \eqref{eq:appD-rho1}, so that we finally obtain
\begin{equation}
    \widetilde{u_i'' u_j''} = \widetilde{u_i'' u_j''}_0 + \widetilde{u_i'' u_j''}_1~,
\end{equation}
where
\begin{equation}
    \widetilde{u_i'' u_j''}_0 = \dfrac{1}{\rho_0}\displaystyle\int \d^3\mathbf{y} ~ \rho_0 \uti{i} \uti{j} K^{\mathbf{x}}
    \label{eq:appD-rey0}
\end{equation}
and
\begin{equation}
    \begin{array}{ll}
        \widetilde{u_i'' u_j''}_1 =& - \dfrac{\widetilde{u_i''u_j''}_0}{\rho_0} \displaystyle\int \d^3\mathbf{y} ~ \rho_0 \xiosci{k} \partial_k K^{\mathbf{x}} \\
        \vspace{0.05cm} \\
        & + \dfrac{1}{\rho_0}\displaystyle\int \d^3\mathbf{y} ~ \rho_0 \xiosci{k} \partial_k \left(\uti{i} \uti{j} K^{\mathbf{x}}\right) \\
        \vspace{0.05cm} \\
        & + \dfrac{1}{\rho_0}\displaystyle\int \d^3\mathbf{y} ~ \rho_0 \uti{i} \uosci{j} K^{\mathbf{x}} \\
        \vspace{0.05cm} \\
        & + \dfrac{1}{\rho_0}\displaystyle\int \d^3\mathbf{y} ~ \rho_0 \uti{j} \uosci{i} K^{\mathbf{x}}~.
    \end{array}
    \label{eq:appD-rey1}
\end{equation}

Additionally, we immediately deduce the linearisation of the turbulent kinetic energy, which corresponds to half the trace of the Reynolds stress tensor
\begin{equation}
    k = k_0 + k_1 = \dfrac{\widetilde{u_i''u_i''}_0}{2} + \dfrac{\widetilde{u_i''u_i''}_1}{2}~.
    \label{eq:appD-k1}
\end{equation}
Likewise, from Eq. \eqref{eq:SPH_epsilon}, we find the linearised turbulent dissipation
\begin{equation}
    \epsilon = \epsilon_0 + \epsilon_1 = \omega_t k_0 + \omega_t k_1~.
    \label{eq:appD-e1}
\end{equation}

\subsubsection{Drift tensor}

The drift tensor $G_{ij}$ being dependent on the mean flow, it also needs to be linearised. We recall that, in its most general form, it can be written as an arbitrary function of the Reynolds stress tensor, the shear tensor and the turbulent dissipation
\begin{equation}
    G_{ij} = G_{ij}\left(\widetilde{u_k'' u_l''}, \partial_k \widetilde{u_l}, \epsilon\right)~,
\end{equation}
and therefore its linearisation reads
\begin{equation}
    G_{ij} = G_{ij,0} + G_{ij,1}~,
\end{equation}
with
\begin{equation}
    G_{ij,0} = G_{ij}\left(\widetilde{u_k'' u_l''}_0, 0, \epsilon_0\right)
    \label{eq:appD-Gij0}
\end{equation}
and
\begin{equation}
    G_{ij,1} = \dfrac{\partial G_{ij}}{\partial \widetilde{u_k'' u_l''}} \widetilde{u_k'' u_l''}_1 + \dfrac{\partial G_{ij}}{\partial (\partial_k\widetilde{u_l})} \left(\partial_k\widetilde{u_l}\right)_1 + \dfrac{\partial G_{ij}}{\partial \epsilon} \epsilon_1~.
    \label{eq:appD-Gij1}
\end{equation}
In Eq. \eqref{eq:appD-Gij0}, $\widetilde{u_k'' u_l''}_0$ is given by Eq. \eqref{eq:appD-rey0} and $\epsilon_0$ by Eq. \eqref{eq:appD-e1}. In Eq. \eqref{eq:appD-Gij1}, the derivatives $\partial G_{ij} / \partial\widetilde{u_k'' u_l''}$, $\partial G_{ij} / \partial(\partial_k\widetilde{u_l})$, and $\partial G_{ij} / \partial \epsilon$ only depend on the functional form of the drift tensor; $\widetilde{u_k'' u_l''}_1$ is given by Eq. \eqref{eq:appD-rey1}; $\left(\partial_k \widetilde{u_l}\right)_1$ is given by Eq. \eqref{eq:appD-shear1}; and $\epsilon_1$ is given by Eq. \eqref{eq:appD-e1}.

\subsection{Linearising the displacement and motion equations\label{app:lin-total}}

Hypotheses (H1) and (H2), in addition to the linearised mean fields computed in the previous section, allow us to write the linearised version of Eqs. \eqref{eq:GLM_eulx_v0} and \eqref{eq:GLM_eulu_v0} as
\begin{equation}
    \dfrac{\partial \xiosci{i}}{\partial t} = \uti{i} + \uosci{i} + \xiosci{j} \dfrac{\partial \uti{i}}{\partial x_j} + \xi_\text{t,j}\dfrac{\partial \uti{i}}{\partial x_j} + \xi_\text{t,j}\dfrac{\partial \uosci{i}}{\partial x_j}
\end{equation}
and
\begin{multline}
    \dfrac{\partial \uosci{i}}{\partial t} + \uti{j} \dfrac{\partial \uti{i}}{\partial x_j} + \uosci{j} \dfrac{\partial \uti{i}}{\partial x_j} + \uti{j} \dfrac{\partial \uosci{i}}{\partial x_j} = \\
    -\dfrac{1}{\rho_0}\dfrac{\partial p_0}{\partial x_i} + \dfrac{\rho_1}{\rho_0^2}\dfrac{\partial p_0}{\partial x_i} - \dfrac{1}{\rho_0}\dfrac{\partial p_1}{\partial x_i} + g_{i,0} + G_{ij,0} \uti{j} \\
    + G_{ij,0} \left(\uosci{j} - \widetilde{u_j}_1\right) + G_{ij,1} \uti{j} + \left[\sqrt{C_0\omega_t k_0} + \dfrac{1}{2}\sqrt{\dfrac{C_0\omega_t}{k_0}}k_1\right] \eta_i~,
    \label{eq:appD-start-uosc}
\end{multline}
where $\rho_0$ and $p_0$ are the equilibrium density and gas pressure, $\rho_1$ is given by Eq. \eqref{eq:appD-rho1}, $p_1$ by Eq. \eqref{eq:appD-p1}, $G_{ij,0}$ by Eq. \eqref{eq:appD-Gij0}, $G_{ij,1}$ by Eq. \eqref{eq:appD-Gij1}, $\widetilde{\mathbf{u}}_1$ by Eq. \eqref{eq:appD-u1}, and $k_0$ and $k_1$ by \eqref{eq:appD-k1}.

Furthermore, we split the right-hand side of Eq. \eqref{eq:appD-start-uosc} three ways: we gather all the terms that do not depend on the oscillatory variables $\uosc$ and $\xiosc$ in a quantity $\mathbf{L}_0$, all the terms that depend on $\xiosc$ and/or $\uosc$ but not on any of the turbulent fields $\bm{\xi_t}$ or $\ut$ in a quantity $\mathbf{L}_1^d$, and all the terms that depend on both the oscillatory variables and the turbulent fields in a quantity $\mathbf{L}_1^s$. This leads us to the following linear equations
\begin{align}
    & \dfrac{\partial \xiosc}{\partial t} - \uosc - (\xiosc \cdot \bm{\nabla})\ut - (\bm{\xi_t} \cdot \bm{\nabla})\uosc = \ut + (\bm{\xi_t} \cdot \bm{\nabla})\ut~, \label{eq:appD-xiosc} \\
    & \dfrac{\partial \uosc}{\partial t} - \mathbf{L}_1^d - \mathbf{L}_1^s = \mathbf{L}_0~,
    \label{eq:appD-uosc}
\end{align}
where
\begin{align}
    L_{1,i}^d &= \left[ \dfrac{1}{\rho_0}\dfrac{\partial p_0}{\partial x_i} - \dfrac{\partial c_0^2}{\partial x_i} \right] \dfrac{1}{\rho_0}\displaystyle\int \d^3\mathbf{y} ~ \rho_0(\mathbf{y}) \left.\left(\xiosci{j} \partial_j K^{\mathbf{x}}\right)\right|_{\mathbf{y},t} \nonumber \\
    & + \dfrac{c_0^2}{\rho_0} \displaystyle\int \d^3\mathbf{y} ~ \rho_0(\mathbf{y}) \left.\left(\xiosci{j} \partial_j \partial_i K^{\mathbf{x}}\right)\right|_{\mathbf{y},t} \nonumber \\
    & + G_{ij,0}\left( \uosci{j} - \dfrac{1}{\rho_0(\mathbf{x})}\displaystyle\int \d^3\mathbf{y} ~ \rho_0(\mathbf{y}) \uosci{j}(\mathbf{y}) K^{\mathbf{x}}(\mathbf{y}) \right)~,
\end{align}
\begin{align}
    L_{1,i}^s &= -\uosci{j} \partial_j \uti{i} - \uti{j} \partial_j \uosci{i} \nonumber \\
    & -G_{ij,0} \dfrac{1}{\rho_0}\displaystyle\int \d^3\mathbf{y} ~ \rho_0(\mathbf{y}) \left.\left( \xiosci{k} \partial_k \left(\uti{j} K^{\mathbf{x}} \right) \right)\right|_{\mathbf{y},t} \nonumber \\
    & + \left(\dfrac{\partial G_{ij}}{\partial \widetilde{u_k'' u_l''}} \widetilde{u_k'' u_l''}_1 + \dfrac{\partial G_{ij}}{\partial (\partial_k\widetilde{u_l})} \partial_k \widetilde{u_l} + \dfrac{\partial G_{ij}}{\partial \epsilon} \omega_t k_1 \right) \uti{j} \nonumber \\
    & + \dfrac{1}{2}\sqrt{\dfrac{C_0 \omega_t}{k_0}} k_1 \eta_i~,
\end{align}
\begin{equation}
    L_{0,i} = -\uti{j}\partial_j \uti{i} - \dfrac{1}{\rho_0} \dfrac{\partial p_0}{\partial x_i} + g_{i,0} + G_{ij,0} \uti{j} + \sqrt{C_0 \omega_t k_0} \eta_i~,
    \label{eq:appD-L0}
\end{equation}
where $c_0^2 \equiv p_0\gamma / \rho_0$ is the equilibrium sound speed squared, and the quantities $\widetilde{u_k'' u_l''}_1$, $\partial_k \widetilde{u_l}_1$, and $k_1$ are given by Eqs. \eqref{eq:appD-rey1}, \eqref{eq:appD-shear1}, and Eq. \eqref{eq:appD-k1} respectively.

\subsection{The forcing term\label{app:0th}}

In Eqs. \eqref{eq:appD-xiosc} and \eqref{eq:appD-uosc} the right-hand side represent inhomogeneous stochastic forcing terms. For the moment we have kept all zeroth order terms (i.e. all the terms that are independent of the wave variables $\xiosc$ and $\uosc$) on these right-hand sides, but $\mathbf{L}_0$ can be rearranged into a more compact form, and some terms will in fact prove negligible. Firstly, let us rewrite the first term on the right-hand side of Eq. \eqref{eq:appD-L0}. Using hypothesis (H4), we write the continuity equation for $\ut$ without any contribution from the oscillatory component
\begin{equation}
    \dfrac{\partial\rho}{\partial t} + \rho\dfrac{\partial \uti{j}}{\partial x_j} + \uti{j}\dfrac{\partial \rho}{\partial x_j}~,
    \label{eq:appD-continuity-total}
\end{equation}
where $\rho$ is the sum of the equilibrium value $\rho_0$ and the turbulent fluctuations of the density $\rho_t$. Building on hypothesis (H3), we neglect $\rho_t$ in Eq. \eqref{eq:appD-continuity-total}, so that
\begin{equation}
    \dfrac{\partial \uti{j}}{\partial x_j} = -\dfrac{\uti{j}}{\rho_0} \dfrac{\partial \rho_0}{\partial x_j}~,
\end{equation}
finally allowing us to write
\begin{align}
    \uti{j} \partial_j \uti{i} &= \partial_j(\uti{j}\uti{i}) - \uti{i}\partial_j\uti{j} \nonumber \\
    & = \partial_j(\uti{j}\uti{i}) - \uti{i}\left(-\dfrac{\uti{j}}{\rho_0}\dfrac{\partial \rho_0}{\partial x_j}\right) \nonumber \\
    &= \dfrac{1}{\rho_0}\dfrac{\partial \rho_0 \uti{j}\uti{i}}{\partial x_j}~.
\end{align}

Secondly, it does not come as a surprise that the non-stochastic part of Eq. \eqref{eq:appD-L0} corresponds to the hydrostatic equilibrium condition. If radiative pressure is neglected, we have
\begin{equation}
    -\dfrac{1}{\rho_0} \dfrac{\partial \rho_0 \overline{\uti{i} \uti{j}}}{\partial x_j} - \dfrac{1}{\rho_0}\dfrac{\partial p_0}{\partial x_i} + g_{i,0} = 0~,
\end{equation}
so that
\begin{equation}
    L_{0,i} = -\dfrac{1}{\rho_0} \dfrac{\partial ( \rho_0\uti{i} \uti{j} - \rho_0 \overline{\uti{i} \uti{j}} )}{\partial x_j} + G_{ij,0} \uti{j} + \sqrt{C_0 \omega_t k_0} \eta_i~.
\end{equation}

It thus becomes clear that the forcing term contains the usual contribution from the fluctuations of the turbulent pressure, a contribution that is linear in $\ut$, and a contribution that is linear in $\bm{\eta}$, and therefore completely uncorrelated in space. Following the discussion from \citet{samadiG01}, we argue that all linear contributions are negligible. The contribution of a linear term to the excitation rate of the modes has an efficiency that is based on the resonance between the lifetime of the large-scale energy-bearing eddies and the period of the modes, which the authors showed was negligible. Naturally, the same argument can be used to neglect the third term as well since it has no coherence in either space or time. The non-linear term, on the other hand, is able to couple different length scales together, and therefore leads to a non-negligible contribution to the excitation rate. Finally, after having filtered out those terms we deemed negligible, we obtain
\begin{equation}
    L_{0,i} = -\dfrac{1}{\rho_0} \dfrac{\partial \left(\rho_0 \uti{i}\uti{j} - \rho_0 \overline{\uti{i} \uti{j}}\right)}{\partial x_j}~.
\end{equation}

The right-hand side of Eq. \eqref{eq:appD-xiosc} can be treated similarly: the term $\ut$ being linear in the turbulent fields, its contribution to mode driving can be neglected, thus only leaving the second term.

\end{appendix}

\end{document}